\documentclass[10pt,preprint]{aastex}
\usepackage{graphicx,apjfonts,emulateapj5,onecolfloat5}

\footnotesize
\newdimen\digitwidth    
\setbox0=\hbox{\rm0}
\digitwidth=\wd0
\catcode`!=\active
\def!{\kern\digitwidth}
\normalsize

\shorttitle{Pulsar rotation measures and Galactic magnetic field}
\shortauthors{Han et al.}
\slugcomment{ApJ accepted -- 2006 Jan.16th}
 
\begin{document}
\twocolumn[

\title{Pulsar rotation measures and the large-scale structure of \\ Galactic magnetic field}
\author{J. L. Han\altaffilmark{1},
        R. N. Manchester\altaffilmark{2},
	A. G. Lyne\altaffilmark{3},
        G. J. Qiao\altaffilmark{4}, and 
        W. van Straten\altaffilmark{5}
	}

\begin{abstract}
The large-scale magnetic field of our Galaxy can be probed in three
dimensions using Faraday rotation of pulsar signals. We report on the
determination of 223 rotation measures from polarization observations 
of relatively distant southern pulsars made using the Parkes radio
telescope. Combined with previously published observations these data give
clear evidence for large-scale counterclockwise fields (viewed from the
north Galactic pole) in the spiral arms interior to the Sun and weaker
evidence for a counterclockwise field in the Perseus arm. However, in
interarm regions, including the Solar neighbourhood, we present evidence
that suggests that large-scale fields are clockwise. We propose that the
large-scale Galactic magnetic field has a bisymmetric structure with
reversals on the boundaries of the spiral arms. Streaming motions associated
with spiral density waves can directly generate such a structure from an
initial inwardly directed radial field. Large-scale fields increase toward
the Galactic Center, with a mean value of about 2~$\mu$G in the Solar
neighbourhood and 4~$\mu$G at a Galactocentric radius of 3 kpc.
\end{abstract}

]

\altaffiltext{1}{National Astronomical Observatories, Chinese Academy of 
	Sciences, Jia 20 DaTun Road, Beijing 100012, China. 
	Email: hjl@bao.ac.cn}
\altaffiltext{2}{Australia Telescope National Facility, CSIRO, PO Box 76,
        Epping, NSW 2121, Australia. Email: rmanches@atnf.csiro.au}
\altaffiltext{3}{University of Manchester, Jodrell Bank Observatory,
	Macclesfield, SK11, 9DL, UK}
\altaffiltext{4}{Department of Astronomy, Peking University (PKU),
	Beijing 100871, China}
\altaffiltext{5}{Center for Gravitational Wave Astronomy, University of
	Texas at Brownsville, TX 78520, USA}

\section{Introduction}

A diffuse magnetic field exists on all scales in our Galaxy. This
field can be detected through observations of Zeeman splitting of
spectral lines, of polarized thermal emission from dust at mm, sub-mm
or infrared wavelengths, of optical starlight polarization due to
anisotropic scattering by magnetically-aligned dust grains, of radio
synchrotron emission, and of Faraday rotation of polarized radio
sources \citep[see][for a review]{hw02}. The first two approaches have
been used to detect respectively the line-of-sight strength and the
transverse orientation of magnetic fields in molecular clouds
\citep[e.g.][]{cru99,ncr+03,fram03}. Starlight polarization can be
used to delineate the orientation of the transverse magnetic field in
the interstellar medium within 2 or 3 kpc of the Sun. Careful analysis
of such data show that the local field is mainly
parallel to the Galactic plane and follows local spiral arms
\citep[e.g.][]{hei96}. Since we live near the edge of the Galactic
disk, we cannot have a face-on view of the global magnetic field
structure in our Galaxy through polarized synchrotron emission, as is
possible for nearby spiral galaxies
\citep[e.g.][]{bbm+96}. Polarization observations of synchrotron
continuum radiation from the Galactic disk \citep[e.g.][]{rfr+02} give the transverse
direction of the field in the emission region and some indication of
its strength.  Large-angular-scale features are seen emerging from the
Galactic disk, for example, the North Polar Spur
\citep[e.g.][]{jfr87,dhjs97,drrf99,rfr+02}, and the vertical filaments
near the Galactic center \citep{hsg+92,dhr+98}. There are also many
small-angular-scale structures resulting from diffuse polarized
emission at different distances which are modified by foreground
Faraday screens \citep{gdm+01,ul02,hkd03a}.

Faraday rotation of linearly polarized radiation from pulsars and
extragalactic radio sources is a powerful probe of the diffuse
magnetic field in the Galaxy
\citep[e.g.,][]{sk80,sf83,ls89,rk89,hq94,hmbb97,id98,fsss01}. Faraday
rotation gives a measure of the line-of-sight component of the
magnetic field. Extragalactic sources have the advantage of large
numbers but pulsars have the advantage of being spread through the
Galaxy at approximately known distances, allowing direct
three-dimensional mapping of the field.  Pulsars also give a
direct estimate of the strength of the field through normalisation by
the dispersion measure (DM). The rotation measure (RM) is defined by
\begin{equation}
\phi = {\rm RM}\; \lambda^2
\end{equation}
where $\phi$ is the position angle in radians of linearly polarised
radiation relative to its infinite-frequency ($\lambda = 0$) value and
$\lambda$ is its wavelength (in m). For a pulsar at distance $D$ (in pc),
the RM (in rad~m$^{-2}$) is given by
\begin{equation}
{\rm RM} = 0.810 \int_{0}^{D} n_e {\bf B} \cdot d{\bf l},
\end{equation}
where $n_e$ is the electron density in cm$^{-3}$, ${\bf B}$ is the vector
magnetic field in $\mu$G and $d {\bf l}$ is an elemental vector along the
line of sight toward us (positive RMs correspond to fields directed toward
us) in pc. With
\begin{equation}
{\rm DM}=\int_{0}^{D} n_e d l,
\end{equation}
we obtain a direct estimate of the field strength weighted by the local free
electron density
\begin{equation}\label{eq_B}
\langle B_{||} \rangle  = \frac{\int_{0}^{D} n_e {\bf B} \cdot d{\bf
l} }{\int_{0}^{D} n_e d l } = 1.232 \;  \frac{\rm RM}{\rm DM}.
\label{eq-B}
\end{equation} 
where RM and DM are in their usual units of rad m$^{-2}$ and cm$^{-3}$ pc and
$B_{||}$ is in $\mu$G.

\citet{man72,man74} was first to systematically measure a number of
pulsar RMs and to use them to investigate the large-scale Galactic magnetic
field. He concluded that the local uniform field is directed toward Galactic
longitude $l\sim90\degr$. \citet{tn80} modeled the magnetic field
configuration to fit the 48 pulsar RMs available that time, and they
confirmed the local field direction and also found evidence for a field
reversal near the Carina--Sagittarius arm. After the large pulsar RM dataset
was published by \citet{hl87}, \citet{ls89} used 185 pulsar RMs to study the
Galactic magnetic field and confirmed the field reversal found by
\citet{tn80}.  \citet{rl94} observed 27 RMs of distant pulsars in the first
Galactic quadrant and provided evidence for a clockwise field (viewed from
the north Galactic pole) near the Crux--Scutum arm. These field directions
were recently re-examined by \citet{wck+04} using revised pulsar distances
from the NE2001 electron density model \citep{cl02} together with their 17
new RMs, finding evidence for several field reversals both exterior and
interior to the Solar circle. \citet{hmq99} observed 54 RMs and tentatively
identified a counterclockwise field near the Norma arm, which was later
confirmed by \citet{hmlq02} using the RM data discussed in this paper. More
recently, \citet{val05} has reanalysed the available pulsar RM data and
interpreted it in terms of an overall clockwise field with a
counterclockwise ring of width $\sim 1$ kpc and radius $\sim 5$ kpc centered
on the Galactic Center.

Pulsar RMs have also been used to study the small-scale random magnetic
fields in the Galaxy. Some pairs of pulsars, which are close in sky position
and have similar DMs, have very different RMs, indicating an irregular field
structure on scales about 100~pc \citep{ls89}. Some of these irregularities
may result from HII regions in the line of sight to the pulsar
\citep{mwkj03}. \citet{rk89} fitted the single-cell-size model for the
residuals of pulsar RMs after the RM contribution of the proposed
large-scale ring-field structure was subtracted and obtained a strength for
the random field $B_r\sim5\mu$G.  \citet{os93} analyzed the difference of
RMs and DMs of pulsar pairs and concluded that $B_r\sim 4-6\mu$G
independent of cell-size in the range of 10 -- 100~pc. In fact the random
fields exist on all scales. \citet{hfm04} have found the power-law
distribution for magnetic field fluctuations as $E_B(k)= C \ (k / {\rm
kpc^{-1}})^{-0.37\pm0.10}$ at scales from $1/k=$ 0.5~kpc to 15~kpc,with $C=
(6.8\pm0.8)\ 10^{-13} {\rm erg \ cm^{-3} \ kpc}$, corresponding to an rms
field of $\sim6\mu$G in the scale range. The interstellar magnetic field is
stronger at smaller scales and may be strongest at the scales of energy
injection by supernova explosions and stellar winds (1 -- 10~pc).

The Parkes multibeam survey has discovered a large number of low-latitude
and relatively distant pulsars \citep{mlc+01,mhl+02,kbm+03,hfs+04},
providing a unique opportunity to probe the diffuse magnetic field in a
substantial fraction of the Galactic disk with much improved spatial
resolution. In addition, improved estimates of pulsar distances are
available from the NE2001 electron density model \citep{cl02}. In this
paper, we adopt a distance of the Sun from the Galactic center of
$R_{\odot}=8.5$~kpc.

We have used the Parkes 64-m telescope of the Australia Telescope
National Facility to observe the polarization properties of 270
pulsars, most of which were discovered in the Parkes Multibeam Pulsar
Survey. After processing, we obtained 223 pulsar RMs which we present
in \S2. All available pulsar RM data have been used to reveal magnetic
field directions along the spiral arms and in interarm regions, as
presented in \S3. The field strength and its Galactocentric radial
dependence are analysed in \S4. Our model for the large-scale Galactic
field is discussed in \S5 and concluding remarks are in \S6.

\section{Pulsar Rotation Measures}
\subsection{Observations and data analysis}

In observation sessions in 1999 December 12--17, 2000 December 14--19
and 2003 February 18--21 we made polarization observations at 20-cm
wavelength of 270 pulsars using the central beam of the 13-beam
multibeam receiver \citep{swb+96} on the Parkes 64-m telescope.  The
receiver is a dual-channel cryogenic system sensitive to orthogonal
linear polarizations with system equivalent flux density of about 29
Jy. Depending on their mean flux density, pulsars were observed for 10
-- 30 min at each of two feed angles, $\pm 45\degr$. For the first two
sessions, signals centred on 1318.5 MHz with a bandwidth of 128 MHz
were processed in the Caltech correlator \citep{nav94}, which gives
128 lags in each of four polarization channels and folds the data
synchronously with the pulsar period in up to 1024 bins per
period. Because of low gain and other instrumental effects, the upper
5 per cent and the lower quarter of the band were given zero weight,
so that the effective bandwidth was about 90MHz. For the 2003 session
a new wideband correlator with a bandwidth of 256 MHz centred at 1375
MHz and $4\times 1024$ lags was used. In all cases, the data were
transformed to the frequency domain, calibrated to give Stokes
parameters, dedispersed to form between 8 and 64 frequency sub-bands
and corrected for parallactic angle variations and ionospheric Faraday
rotation -- see \citet{nms+97} for details. The ionospheric RM was
typically between $-1$ and $-5$ rad m$^{-2}$ with a largely diurnal
variation. The 2003 observations were processed using the {\sc
  psrchive} software package \citep{hvm04}.

In off-line analysis, corresponding sub-bands from feed-angle pairs
were added after normalization by the area of the Stokes I
profile. Summing of the orthogonal feed-angle pairs eliminates most of
the effects of polarisation cross-coupling in the feed.  To determine
the rotation measure, we searched for a peak in the total linearly
polarized intensity $L = (Q^2 + U^2)^{1/2}$ obtained by summing in
frequency using a set of trial RMs, normally in the range of $\pm$2000
rad~m$^2$ with a step of about 20 rad~m$^2$. Then, using the RM value
corresponding to the peak, the data were summed to form upper and
lower band profiles.  Finally, the best estimate of the RM was then
obtained by taking the weighted mean position-angle difference across
the profile between the two bands, with the weight inversely
proportional to the square of the error in position angle difference
for each pulse phase bin. This procedure reduces the significance of
the problem related to transitions between orthogonal modes discussed
by \citet{rbr+04}.

\begin{figure*} 
\includegraphics[angle=270,width=180mm]{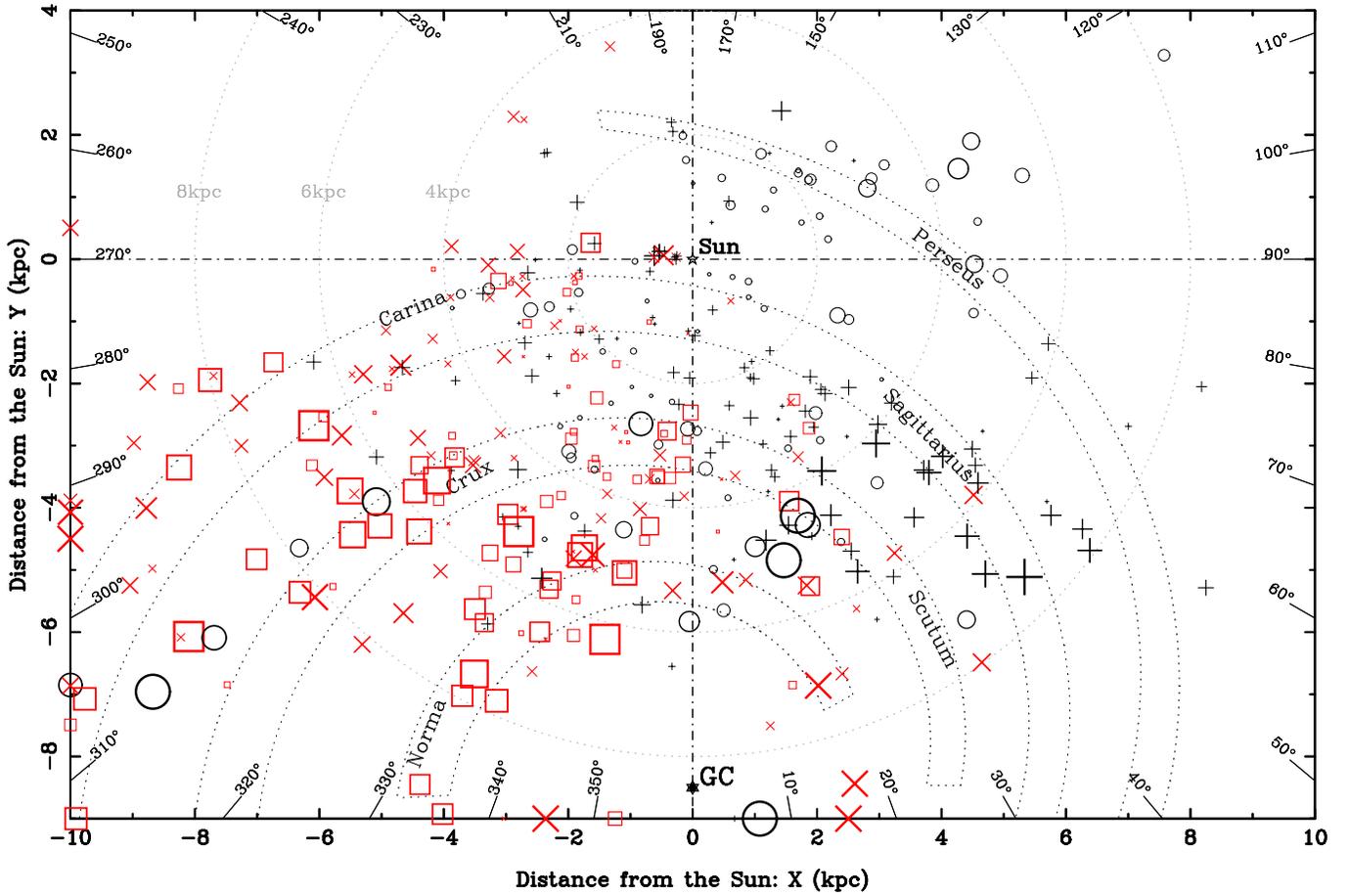}
\caption{The RM distribution of 388 pulsars with $|b|<8\degr$,
projected onto the Galactic Plane. The linear sizes of the symbols are
proportional to the square root of the RM values, with limits of 9 and
900 rad m$^{-2}$. The crosses (+ and $\times$) represent positive RMs,
and the open circles and squares represent negative RMs. New
measurements are indicated by $\times$ and squares.  Approximate
locations of four spiral arms are indicated (see text). Distances to the pulsars
are based on the NE2001 model for the Galactic electron density
distribution \citep{cl02}.  Pulsars with estimated distances too large
to show are plotted on the edge of the figure. }
\label{rm-xy}
\end{figure*}
\begin{figure*} 
\includegraphics[angle=270,width=180mm]{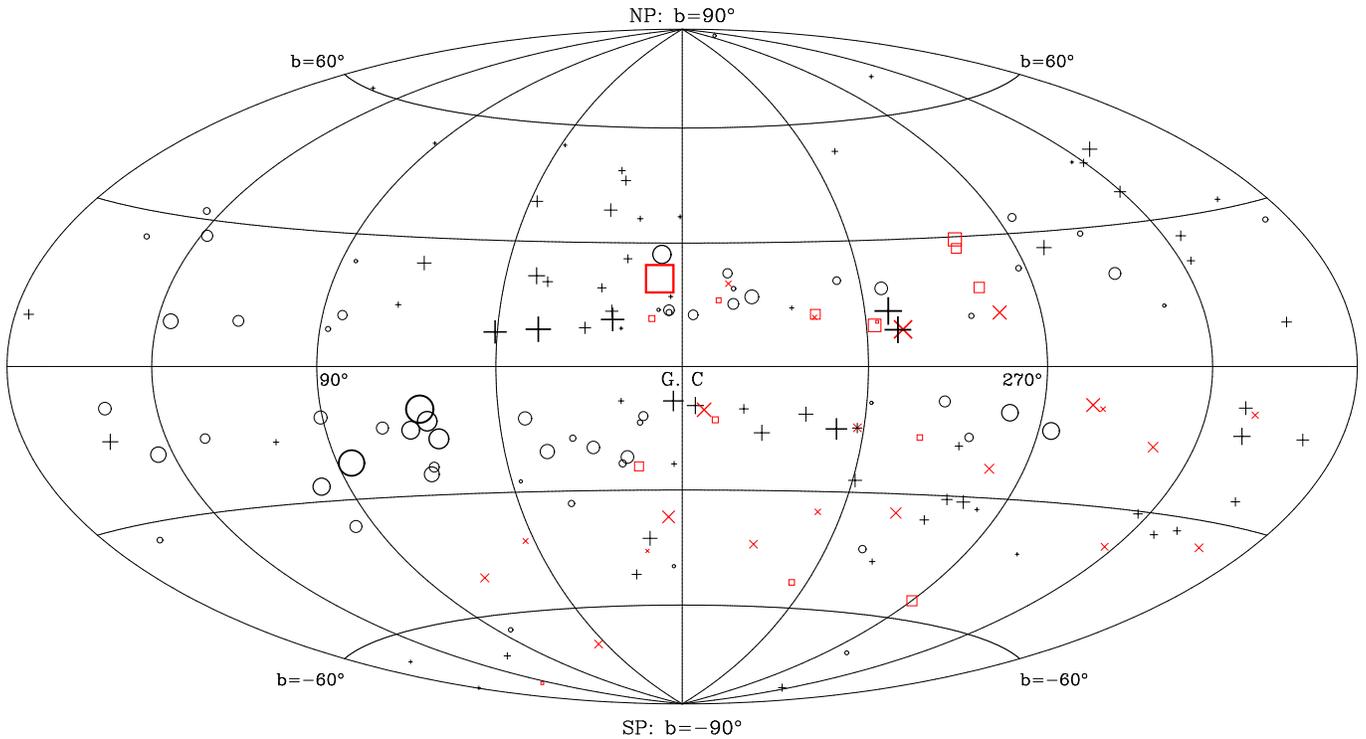}
\caption{The Galactic distribution of RMs for 166 pulsars with
$|b|>8\degr$, including 36 new measurements. 
The linear sizes of the symbols are proportional to
the square root of the RM values, with limits of 2.7 and 270 rad
m$^{-2}$. See Figure~\ref{rm-xy} for an explanation of the symbols.}
\label{rm-lb}
\end{figure*}
\begin{figure*} 
\includegraphics[angle=270,width=185mm]{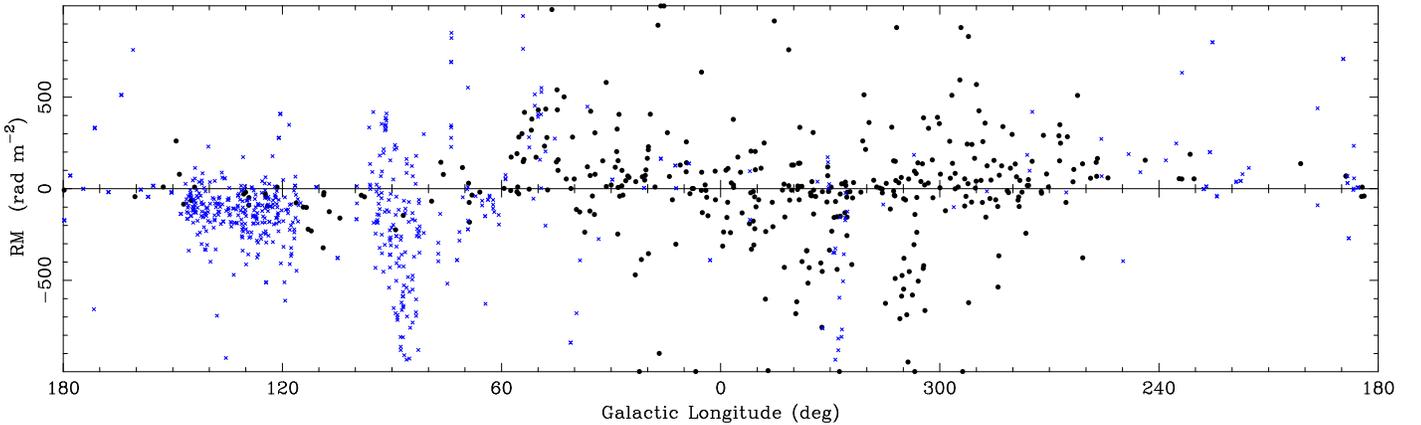}
\caption{Distribution of RMs with Galactic longitude for pulsars (dots) and
extra-galactic radio sources (crosses) with Galactic latitudes 
$|b|<8\degr$. The RMs for extra-galactic sources are from
\citet{ccsk92}, \citet{gdm+01} and \citet{btj03}.}
\label{gl-rm}
\end{figure*}
\subsection{The RM database}
Rotation measures for 223 pulsars from the Parkes observations are
listed in Table~1. The first two columns give the pulsar name based on
J2000 coordinates and, where assigned, the name based on B1950
coordinates. Columns 3 to 6 give the pulsar period, DM and Galactic
longitude and latitude. The pulsar distance given in column 7 is based
on the pulsar DM and the NE2001 model for the Galactic distribution of
interstellar electrons \citep{cl02}. For pulsars at distances greater
than 5~kpc, \citet{cl02} estimate that the derived distances have
uncertainties of 10\% in most of the fourth Galactic quadrant, but up
to 20\% near $l\sim 270\degr$. The measured RM and its estimated
standard error are given in the next two columns and the final column
gives the observing session for that pulsar.

Of the 223 RMs in Table 1, nine are RMs of millisecond pulsars based
on the present data set \citep{mh04} which are included here for
completeness. Excluding these, there are 27 pulsars in Table~1 for
which previously published RM data exist. These 27 pulsars are listed
in Table~2 showing that most of new values are consistent with, and
sometimes better than, previous measurements. Small significant
differences may be real in some cases, reflecting a changing RM. In
some previous measurements, ionospheric RMs have not been taken into
account, also leading to differences of a few rad m$^{-2}$.

Five sources deserve particular comment.  PSR J1116$-$4122
(B1114$-$41) has a very narrow and weakly polarized pulse. Both
\citet{vdhm97} and \citet{hmq99} failed to get a good signal-to-noise
ratio for the linear polarization to allow a reliable determination of the
RM. Our integration time is longer and we believe our new measurement is
reliable. 
The rapid position-angle sweep occurring in the narrow pulse of PSR
J1946$-$2913 was not well resolved by \citet{hmq99}.  We observed with the
full 1024-bin resolution and obtained a much better quality polarization
profile and hence an improved value for the RM.
The RM of PSR J2330$-$2005 seems to have steadily increased with
time, from 9.5$\pm$0.2 rad~m$^{-2}$ \citep{hmm81} to 16$\pm$3
\citep{hl87} to 30$\pm$7 rad~m$^{-2}$ (this paper). This suggests that the
line of sight is traversing a very compact magneto-ionic region. 
Our measured RM value for PSR J1757$-$2421 (B1754$-$24) is
$-9\pm9$~rad~m$^{-2}$, compared with the value of $+153\pm12$
rad~m$^{-2}$ given by \citet{hl87}.  However, those authors said that
the true RM for this pulsar could be smaller by
153 rad~m$^{-2}$ because of an unresolved ambiguity in their
observations.  Our measurement resolves this ambiguity and indicates
that their value should be modified to $0\pm12$ rad~m$^{-2}$, bringing
the two values into good agreement. 
Our measurements suggest that the RM for PSR J1141$-$6545 published in
the discovery paper \citep{klm+00} has an incorrect sign.

We have used the
\anchor{http://www.atnf.csiro.au/research/pulsar/psrcat}{ATNF Pulsar
  Catalogue}\footnote{See
  http://www.atnf.csiro.au/research/pulsar/psrcat} \citep{mhth05} to
obtain a total of 367 previously published RMs, principally from the
major studies by \citet{hl87,rl94,qmlg95,vdhm97,hmq99,cmk01,mwkj03}
and \citet{wck+04}. Adopting the most reliable value for pulsars with
multiple measurements, we obtain RMs for a total of 554 pulsars. Most
of these have been corrected for the ionospheric RM but, in any case,
these corrections are small and will not significantly affect the
analysis below for the magnetic fields in the Galactic disk.

The RM distribution projected onto the Galactic plane for low-latitude
pulsars ($|b|<8\degr$) is shown in Figure~\ref{rm-xy}.
Figure~\ref{rm-lb} shows the distribution projected onto the celestial
sphere for high-latitude pulsars ($|b|>8\degr$), including 36 new
measurements, which may be used to probe the magnetic fields in the
Galactic halo \citep[e.g.][]{hmbb97}. Pulsars with measured RMs now
cover about one third of the Galactic disk; our new data very
significantly improve the coverage in the fourth Galactic
quadrant. Many pulsars in the inner Galaxy are close to or beyond
tangential points, which enables us to begin to distinguish fields in
the arm and interarm regions.

Radio polarisation surveys of the Galactic plane
\citep[e.g.,][]{gdm+01,btj03} provide many RMs for low-latitude
extragalactic sources. These sample the entire path through the Galaxy
and are especially useful for diagnosing the magnetic field in the
outer Galaxy where there are few known pulsars. Figure~\ref{gl-rm}
shows the distribution in Galactic longitude of RMs for both pulsars
and low-latitude extra-galactic sources.

\section{Large-scale structure of magnetic fields in the Galactic disk}
\label{sect_Bstruct}
Figure~\ref{rm-xy} clearly shows that there is large-scale order in
the distribution of Galactic magnetic fields. Rotation measures are
predominantly positive in the first Galactic quadrant (longitudes
$0\degr - 90\degr$) and negative in the fourth quadrant ($270\degr -
360\degr$), implying an overall counterclockwise field in the
Galaxy. As we discuss further below, closer examination of the data
shows that the field direction is reversed in certain regions, mostly
between the spiral arms. For these large-scale fields, we can assume
that, because of stretching due to differential Galactic rotation, the
azimuthal component of the field, $B_{\phi}$, dominates over both the
radial component, $B_{r}$, and the vertical component, $B_z$
\citep{hmq99}.  Local bubbles which cover a significant solid angle on
the sky \citep{val93a} can result in an offset in the RM values for
pulsars lying behind them; such offsets don't significantly affect our
analysis. Although individual pulsars may be affected by
incorrect distances or HII regions along the line of sight, trends
common to many pulsars average over these effects and give a reliable
measure of the large-scale field.

As discussed by \citet{ls89}, \citet{hmlq02} and \citet{wck+04},
measuring the gradient of RM with distance or DM is a powerful method
of determining both the direction and magnitude of the local
large-scale field in particular regions of the Galaxy. Field strengths
in different regions of the Galaxy can be estimated from the slope of
trends in plots of RM versus DM since, from Equation~\ref{eq_B}, we
have
\begin{equation}
\langle B_{||}\rangle_{d1-d0} = 1.232 \Delta{\rm RM}/\Delta{\rm DM}
\label{delta_rm_dm}
\end{equation}
where $\langle B_{||}\rangle_{d1-d0}$ is the mean line-of-sight field
component in $\mu$G between distances $d0$ and $d1$, $\Delta{\rm RM} =
{\rm RM}_{d1} - {\rm RM}_{d0}$ and $\Delta{\rm DM} ={\rm DM}_{d1} - {\rm DM}_{d0}$.

The locations of spiral arms in our Galaxy is a subject of much debate
\citep[see][for good summaries of the uncertainties]{bur76,rus03}. Based on
observed tangent points (cf. Figure~\ref{B_GL}) and locations of giant
molecular clouds, HII regions and star-formation complexes
\citep{gcbt88,sr89,bact89,dht01,gg76,dwbw80,ch87a,rus03} we have
adopted locations for the major spiral arms as shown in
Figure~\ref{rm-xy}. These arms are approximately equiangular spirals
with pitch angles of between $-10\degr$ and $-12\degr$. While the
locations (or even existence) of these arms is quite uncertain,
especially in directions toward the Galactic Center, most of the
discussion in this paper is based on regions around the tangential
points which are reasonably well determined. Uncertainties in pulsar
distances also have least effect in these directions.

\begin{figure}[bht]
\includegraphics[angle=270,width=86mm]{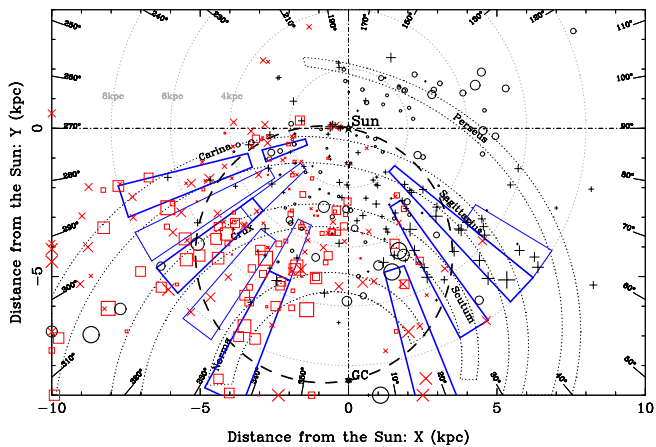}
\caption{Distribution of pulsars with known RMs and $|b|<8\degr$ (as in
Figure~\ref{rm-xy}) with boxes showing the regions near tangential points of
spiral arms and in interarm regions where values of $\langle B_{||}\rangle$
have been estimated using Equation~\ref{delta_rm_dm} in
Fig.~\ref{norma_rm_dm} to \ref{inter_rm_dm} and \ref{pers_rm_dm}. Interarm
boxes are marked with thin lines.  The dashed circle is the locus of
tangential points for equiangular spirals of pitch angle $-11\degr$.}
\label{xy_tgt}
\end{figure}

We have applied Equation~\ref{delta_rm_dm} to regions near the tangential
points of the spiral arms as marked on Figure~\ref{xy_tgt} where the
large-scale field and the lines of sight have maximum projections.  For
regions around Galactic longitude $l\sim0\deg$ (and also $180\deg$), the
large-scale field is nearly perpendicular to our line of sight and hence not
measurable using Faraday rotation.  We have taken all pulsars with RMs
having errors of less than 25 rad m$^{-2}$ with $|b|<8\degr$ and lying
within $4\degr$ of the central longitude and within the distance ranges
indicated on Fig.~\ref{xy_tgt}. The DM range corresponding to the adopted
distance range was computed using the NE2001 model. We have determined
RM--DM slopes using the robust straight-line fit method \citep[see Sect.15.7
in ][]{ptvf92}, which minimizes the effects of anomalous outliers in the
plots due, for example, to HII regions \citep{mwkj03}, to give the mean
field $\langle B_{||}\rangle$ within the region. The uncertainty of $\langle
B_{||}\rangle$ is determined from the absolute mean deviation of RMs from
the fitted line together with the average of DMs of pulsars in the
region. We reject any fits if the uncertainty\footnote{The uncertainty of
the fit reflects the mean deviation of the data which is dominated by random
magnetic fields in the sample region. To judge the direction (i.e., sign) of
the large-scale field, a 2$\sigma$ significance should be adequate; all our
results below are in fact greater than 3$\sigma$.}, $\sigma$, is greater than
1~$\mu$G or if $\langle B_{||}\rangle$ has a significance of less than
2$\sigma$.  Similar fits have been made to interarm regions as indicated in
Fig.~\ref{xy_tgt}. For completeness, we briefly review previously published
results for the large-scale magnetic field in the local region and near and
beyond the Perseus arm.

\begin{figure}[bht] %
\includegraphics[angle=270,width=86mm]{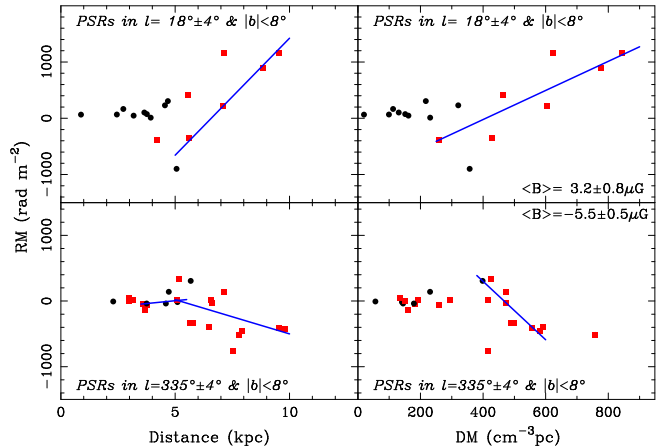}
\caption{Dependence of RM with distance and DM for pulsars lying in
directions passing through the Norma arm ($l=18\degr \pm 4\degr$,
$l=335\degr \pm 4\degr$).  In this and subsequent
figures of this type, RMs from this paper are shown as squares and
previous measurements as dots, and the lines are robust straight-line fits
to RMs for pulsars lying within the defined regions (see text). 
The corresponding mean field from Equation~\ref{delta_rm_dm} and its
statistical uncertainty are shown on the plots.}
\label{norma_rm_dm}
\end{figure}
\subsection{The Norma spiral arm}
The large-scale field in the Norma arm, the inner-most identified arm in the
Galaxy, has been discussed in detail by \citet{hmlq02} based on the present
dataset but using distances based on the \citet{tc93} electron density
model. Here we analyse the data with distances based on the NE2001 electron
density model. Figure~\ref{norma_rm_dm} shows RMs as a function of distance
and DM for low-latitude pulsars lying near $l=18\degr$ and $l=335\degr$. The
fitted lines for distances between 6 and 10 kpc are within the Norma arm
(see Figure~\ref{xy_tgt}).  The line between 3 and 5 kpc at $l=335\degr$
corresponds to the adjacent interarm region and will be discussed in
\S\ref{subsectinter}.  It is clear that the conclusions of \citet{hmlq02}
are maintained. The RM gradient for pulsars between 6 and 10 kpc from the
Sun is positive for $l=18\degr$ and negative for $l=335\degr$ showing that
the magnetic field is coherently counterclockwise along the arm across both
quadrants. Even more negative RMs for extragalactic sources around
$l=330\degr$ \citep[see Figure~\ref{gl-rm}, data from][]{gdm+01} indicate
that the counterclockwise field in the Norma arm is maintained beyond 10~kpc
from the Sun in this direction.

\begin{figure}[bht] %
\includegraphics[angle=270,width=86mm]{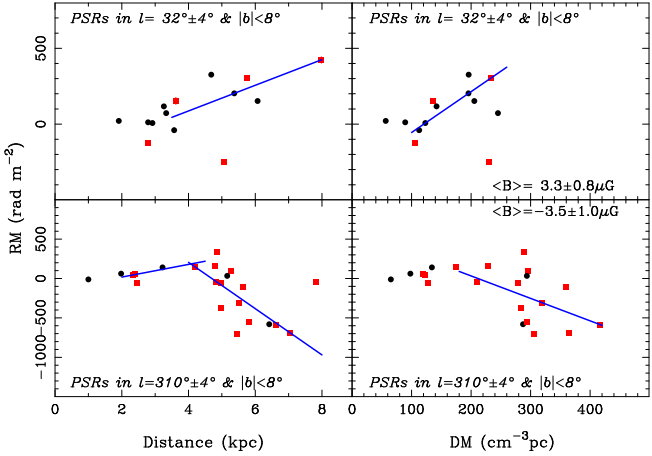}
\caption{Dependence of RM with distance and DM for pulsars lying in
directions passing through the Scutum ($l=32\degr$) 
and Crux ($l=310\degr$) spiral arms. The short fitted line at
$l=310\degr$ is for the interarm region between the Crux and Carina arms. }
\label{CS_rm_dm}
\end{figure}

\subsection{The Crux--Scutum arm} 
The present observations have provided a large number of new RMs in the
regions near the tangential point of the Crux arm (Figure~\ref{rm-xy}). In
this arm, near $l=310\degr$, the RMs are mostly negative, whereas those
around $l=32\degr$, passing along the Scutum arm, are mostly positive,
clearly indicating a counterclockwise field in the whole of the Crux--Scutum
arm. The variation of RM with distance and DM is shown for pulsars in these
directions in Figure~\ref{CS_rm_dm}. For $l=310\degr$ there is a clear
decrease in RM with increasing distance through the Crux arm, indicating a
large-scale field directed away from the Sun. There is a corresponding
increase in RM for pulsars more distant than 3 kpc lying along the Scutum
arm. This tendency is also seen in the data of \citet{wck+04} (their Figure
8). These results are then consistent with a counterclockwise large-scale
field through the whole of the Crux--Scutum arm.

\subsection{The Carina--Sagittarius arm}\label{car-sag}
\begin{figure}[bht] %
\includegraphics[angle=270,width=86mm]{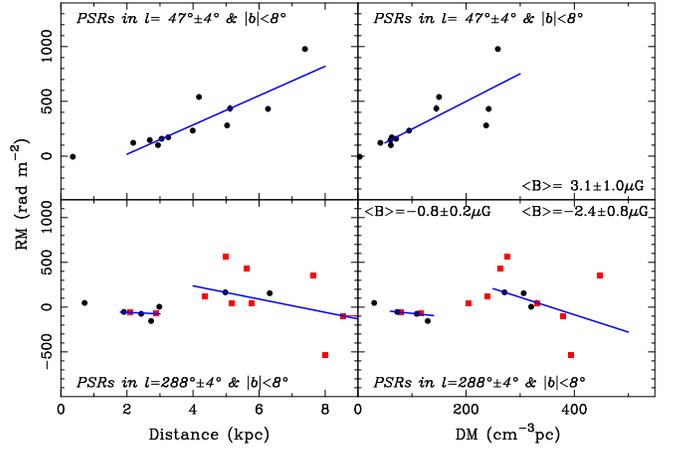}
\caption{Dependence of RM with distance and DM for pulsars lying in
directions passing through the Sagittarius
($l=47\degr$) and  Carina ($l=288\degr$) spiral arms. }
\label{Car_Sag_rm_dm}
\end{figure}

It is clear from Figure~\ref{rm-xy} that the Sagittarius arm is dominated by
positive RMs which increase with distance, indicating a counterclockwise
field in this arm. This is also shown by Figure~\ref{Car_Sag_rm_dm} where
there is a clear trend for RMs of pulsars at longitudes near $47\degr$ in
the distance range 2 -- 7 kpc to increase with distance (cf.
Fig.6 of \citet{wck+04}.) In this case, the adopted tangential
longitude is somewhat inside that for an equiangular spiral
(cf. Figure~\ref{xy_tgt}) but it is consistent with the arm location
adopted by \citet{cl02}. The evidence for a
counterclockwise field in this arm has been discussed by many previous
authors \citep[e.g.][]{ls89,hq94,id98,wck+04}.

The Carina arm is evidently more
complicated. Figure~\ref{Car_Sag_rm_dm} shows a plot of RM versus
distance and DM for pulsars within $4\degr$ of $l=288\degr$. Between 1
and 3~kpc from the Sun, RMs in this direction are increasingly negative,
consistent with a continuation of the counterclockwise field from the
Sagittarius arm \citep[see also][]{fsss01}. However, as
Figure~\ref{rm-xy} shows, the newly determined RMs for distant pulsars
through the Carina arm are unexpectedly positive. At distances greater
than 4 kpc, there is a large scatter in the RM versus distance and DM
plots (Figure~\ref{Car_Sag_rm_dm}) but a fit to the data again shows a
negative slope indicating a counterclockwise field in the outer parts
of this arm. The offset of about +500~rad~m$^{-2}$ between
the two lines indicates a region of reversed field between 3 and 5
kpc from the Sun. The origin of this ``Carina anomaly'' is not clear,
but there are many large HII regions in this region which could have a
significant influence on the RMs of pulsars lying behind them. This
anomalous region may also account for the group of large positive RMs
for very distant pulsars with longitude between $290\degr$ and $295\degr$
(Figure~\ref{rm-xy}).

\subsection{Interarm regions in the inner Galaxy }
\label{subsectinter}
In the first Galactic quadrant, it is difficult to separate arm and
interarm contributions because of the small separation of the arms. A
group of very negative RMs for pulsars between longitudes $15\degr$
and $20\degr$ and lying just beyond the Scutum arm
(Figure~\ref{rm-xy}) was taken by \citet{rl94} as evidence for a
clockwise field near or beyond the Scutum arm, a result adopted by
\citet{hmq99} and \citet{wck+04}.

\begin{figure}[bht]
\includegraphics[angle=270,width=86mm]{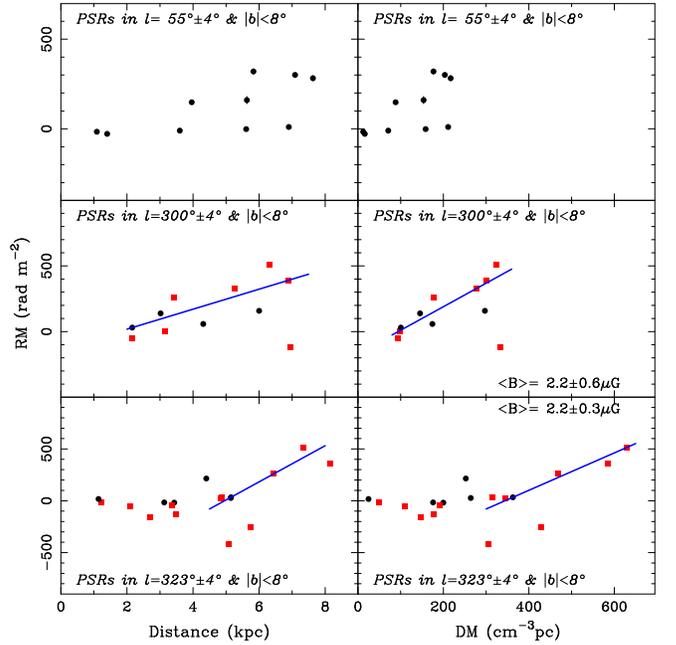}
\caption{Dependence of RM with distance and DM for pulsars lying in
interarm regions for directions around $l=55\degr$ (between the
Sagittarius and Perseus arms), around $l=300\degr$ (between the Carina
and Crux arms), and around $l=323\degr$ (between the Crux and Norma
arms). }
\label{inter_rm_dm}
\end{figure}

Figure~\ref{inter_rm_dm} shows RM plots for several interarm
regions. Three postive RMs for pulsars around $l\sim55\degr$ at about
7.5~kpc (Figure~\ref{rm-xy}) hint at a counter-clockwise field between
the Sagittarius arm and the pulsars, though one of them (PSR
J1927+1852) has a large RM uncertainty ($417\pm70$) and been omitted
in Figure~\ref{inter_rm_dm}. Obviously, this result is of very low
statistical significance.

The situation is better in the fourth quadrant where we have a large
number of new RMs and there is a larger separation of the arms making
distance uncertainties less important. Figure~\ref{norma_rm_dm} shows
that the RM is increasing between 3.5 and 5.5~kpc for $l=335\degr$,
suggesting a clockwise magnetic field between the Crux and Norma
arms. Similarly, an increasing RM between 2 and 4~kpc in
Figure~\ref{CS_rm_dm} indicates the clockwise field between the Carina
and Crux arms.  Figure~\ref{inter_rm_dm} shows the distance and DM
dependence of RMs of more distant pulsars in the interarm regions
between the Carina, Crux and Norma arms. For $l=300\degr$ there is a
clear tendency for increasing RM with distance between 3 and 7 kpc,
suggesting a clockwise field in the region between the Carina and Crux
arms. Similarly, RMs for pulsars lying between 5 and 8 kpc in the
direction $l=323\degr$ show a positive trend, increasing from very
negative values at 5 kpc to positive values at 8 kpc. This again
suggests a clockwise field in this interarm region.

\subsection{The local region}
\label{subsectloc}
\begin{figure}[bht] 
\includegraphics[angle=270,width=86mm]{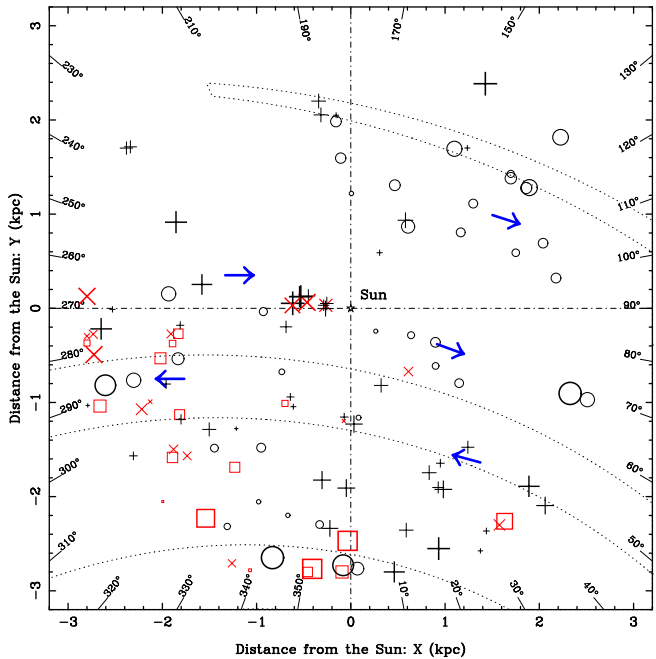} 
\caption{The RM distribution for pulsars with $|b| < 8\degr$ and
distance projected on the Galactic plane less than 3 kpc.  See
Figure~\ref{rm-xy} for an explanation of the symbols. Plotted field
directions are based on both previous studies (see text) and RM--DM
plots given in this section (Figs~\ref{Car_Sag_rm_dm} and \ref{pers_rm_dm}).}
\label{rm_loc} 
\end{figure}

Figure~\ref{rm_loc} gives an expanded view of the the RM distribution
in the local region, i.e. within $\sim 3$~kpc of the Sun.  The
large-scale field in the local interarm region, i.e. between the
Carina--Sagittarius arm and the Perseus arm, has long been known to
have a clockwise direction \citep[e.g.][]{man74}, with mostly negative
RMs for longitudes within $30\degr$ of $l=90\degr$ and mostly positive
for longitudes within $30\degr$ of $l=270\degr$. A few new RM
measurements around $l=270\degr$ reinforce this conclusion. 

As discussed above, pulsars in the Carina and Sagittarius arms have
different RM signs to those in the interarm region, clearly indicating
a reversal of the large-scale field between the arm and interarm
regions.  \citet{hq94} and \citet{id98} have modelled the field in
this region as a bisymmetric spiral after taking into account local
anomalies resulting from bubbles and HII regions \citep{val93a}. The
pitch angle of the spiral field is $-8\degr\pm2\degr$, close to the
value of spiral arms, and its strength is about 1.8$\pm0.3\mu$G
\citep{han01}. Similar conclusions were reached by \citet{fsss01} but
they derived a steeper pitch angle, $\sim -14\degr$.

\subsection{The outer Galaxy}
The direction of the magnetic field in the Perseus arm cannot be 
established with any certainty. Rotation measures for distant pulsars 
and extragalactic sources are strongly negative between longitudes of 
$80\degr$ and $150\degr$ (Figures~\ref{rm-xy} and \ref{gl-rm}). This 
has been used by many authors \citep[e.g.][]{bt01,btwm03,mwkj03} to 
argue for an absence of any reversals in the outer part of the Galaxy, 
through and beyond the Perseus arm. In any case, it seems clear that 
the interarm fields both between the Sagitarius and Perseus arms and 
beyond the Perseus arm are predominantly clockwise. 

\begin{figure}[bht] 
\includegraphics[angle=270,width=86mm]{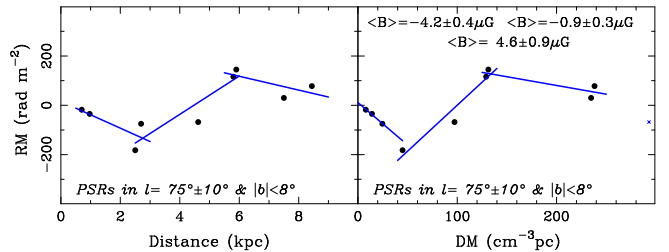} 
\caption{Dependence of RM with distance and DM for pulsars lying within
$10\degr$ of $l=75\degr$.}
\label{pers_rm_dm} 
\end{figure}

Few pulsars are known in the outer parts of the Galaxy, making
determination of the field structure difficult and uncertain. Also,
for longitudes between roughly $130\degr$ and $210\degr$ our line of
sight is nearly perpendicular to the spiral structure, so RMs are not
sensitive to the dominant longitudinal or spiral field component. A
group of four positive RMs around $l\sim70\degr$ to $80\degr$ were
used by \citet{hmq99} and \citet{wck+04} to argue for a
counterclockwise magnetic field in the Perseus arm, reversed from the
local clockwise field. Based on extra-galactic RMs, \citet{fsss01}
also concluded that the field in the Perseus arm is
counterclockwise. However, as Figure~\ref{pers_rm_dm} shows, there is
a weak but apparently significant decrease of RM with increasing DM beyond 
$\sim6$~kpc, indicating a clockwise field exterior to the Perseus arm.
Also, the predominantly
negative RMs of extragalactic radio sources around $l=75\degr$ (see
Figure~\ref{gl-rm}), in contrast to postive RMs of pulsars in or near the
Perseus arm, suggest that magnetic fields exterior to the Perseus arm
are clockwise. The pulsar distance scale is quite uncertain in this
region and it is possible that the four pulsars actually lie (just)
beyond the Perseus arm.

\begin{figure}[thb] %
\includegraphics[angle=270,width=85mm]{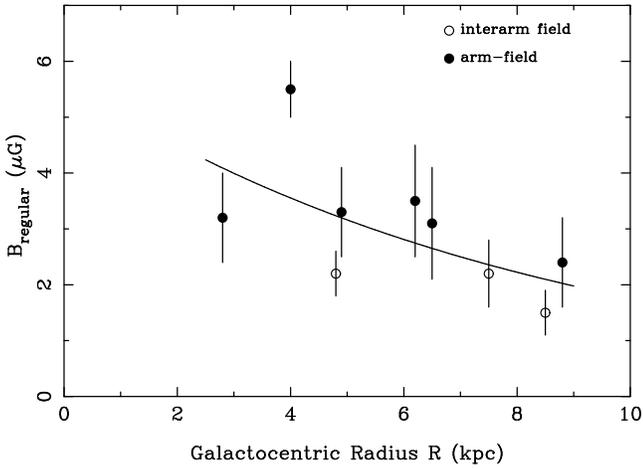}
\caption{Dependence of the strength of the large-scale regular field with
the Galactocentric radius. Filled dots are for arm regions and small open
circles are for interarm regions. The curved line is a fit of an exponential
model -- see text.}
\label{B_R}
\end{figure}

\section{Magnetic field strength}
Various authors have suggested that the large-scale magnetic fields in the
Galactic disk are stronger towards the Galactic Center
\citep[e.g.][]{sf83,rl94,hei96b}. Such a radial variation has been assumed
in modelling of the Galactic synchrotron emission \citep[Berkhuijsen,
unpublished but cited in][]{bbm+96} and the Galactic $\gamma-$ray background
by \citet{smr00}. In the local region, measurements of the mean field
strength give values of $1.5\pm0.4\;\mu$G \citep{hq94,id98}, whereas
\citet{hmlq02} find a value of $4.4\pm0.9\;\mu$G in the Norma arm. With the
much extended pulsar RM data now covering about one third of the Galactic
disk, we are better able to investigate the dependence of field strength on
Galactocentric radius.

Taking the field strengths for directions in the inner Galaxy and the
local field strength from \citet{hq94}, and correcting for the angle
between the assumed dominant azimuthal field direction and the line of
sight, we obtain the field strengths shown in Figure~\ref{B_R} as a
function of Galactocentric radius.  Although uncertainties are large,
there are clear tendencies for fields to be stronger at smaller
Galactocentric radii and weaker in interarm regions.

To parameterize the radial variation, we tried fitting different
functions to the data: a constant \citep{rk89,hq94}, a $1/R$-function
\citep{sf83}, a linear gradient and an exponential function
\citep{smr00}. The exponential function not only gives the smallest
$\chi^2$ value but also avoids the singularity at $R=0$ (for $1/R$)
and unphysical values at large R (for the linear gradient). The fitted
function shown in Figure~\ref{B_R} is
\begin{equation}
B_{\rm reg}(R) = 
B_0 \; \exp \left[ \frac{-(R-R_{\odot})} {R_{\rm B}} \right]
\label{BR}
\end{equation}
with the strength of the large-scale or regular field at the Sun,
$B_0=2.1\pm0.3$ $\mu$G and the scale radius $R_{\rm B}=8.5\pm4.7$ kpc.

\section{Global structure of magnetic fields in the Galactic disk}
\begin{figure}[thb] %
\includegraphics[angle=270,width=89mm]{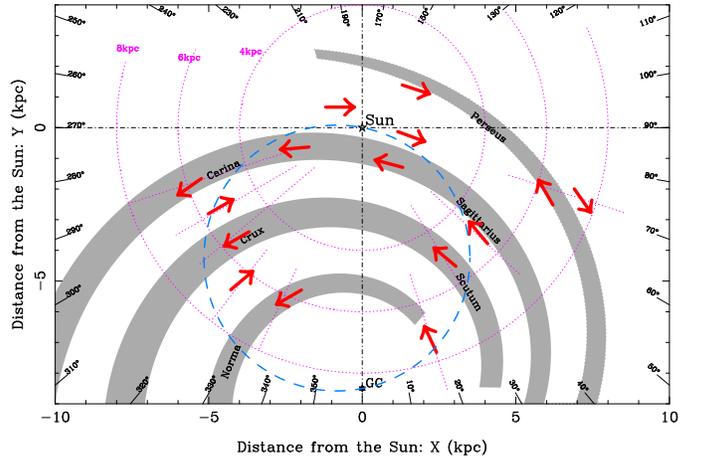}
\caption{Global pattern of magnetic field directions inferred from
RM--DM fits to the pulsar data and assuming an overall spiral pattern for
the large-scale field.  Field directions in the local region ($<$3kpc
from the Sun) and in the Perseus arm were taken from previous studies
\citep[e.g.][]{hq94,id98} (see text). The dashed circle is the locus
of tangential points for equiangular spirals of pitch angle
$-11\degr$. }
\label{global_B}
\end{figure}

Although there remain considerable uncertainties in several regions, a
striking pattern emerges from the above discussion: large-scale magnetic
fields in spiral arms are counterclockwise (viewed from the north) but in
the interarm regions the fields are clockwise. Figure~\ref{global_B}
summarizes the evidence for this bisymmetric global pattern, which is mainly
based on the field directions near the tangential points derived in
\S\ref{sect_Bstruct}, for this bisymmetric global pattern. These data are
relatively insensitive to uncertainties in the pulsar distance scale or
errors in distances to individual pulsars.

\begin{figure}[thb]  
\includegraphics[angle=270,width=85mm]{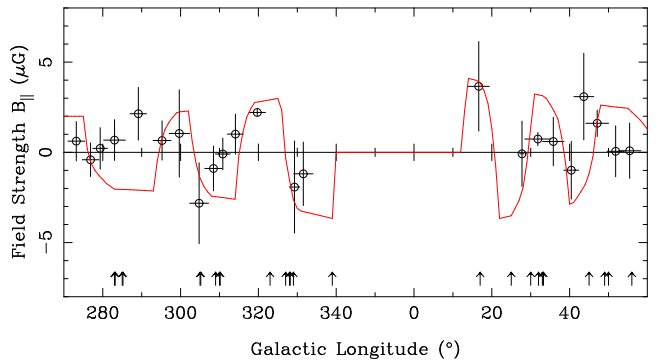}
\caption{Mean line-of-sight field strengths derived from RM vs DM
gradients for pulsars lying near the tangent points of an equiangular
spiral pattern of $-11\degr$ pitch angle as a function of Galactic
longitude. Points are plotted at the mean longitude of the pulsars
lying within a defined region and the crosses represent the rms
scatter of longitude and mean field.   Tangent points for spiral arms
in the inner Galaxy based on observational data
\citep{gcbt88,sr89,bact89,dht01,gg76,dwbw80,ch87a,eg99,dri00,rus03}
are marked by small vertical arrows.
The variation of the mean
tangential field expected for a simplified bisymmetric global model in which
fields are counterclockwise within spiral arms and clockwise in
interarm regions is shown by the solid line. See the text for more
details.}
\label{B_GL}
\end{figure}

To further quantify this evidence, we have used Equation~\ref{delta_rm_dm}
to compute the mean line-of-sight field strength in regions tangential to an
equiangular spiral pattern of pitch angle $-11\degr$. The locus of these
tangential points is shown in Figure~\ref{global_B}. At 4-degree intervals
of longitude, the RM versus DM dependence was determined by a least-squares
fit of a line to data for pulsars with $|b|<8\degr$ lying within a box of
longitude width $8\degr$ centered on the tangential point. The ends of the
box were defined by the points at which the spiral through the tangental
point reached a longitude that is $4\degr$ from the longitude of the
tangental point, typically 1 -- 2 kpc from the tangential point.  RMs with
uncertainties greater than 30 rad m$^{-2}$ were omitted from the
fits. Figure~\ref{B_GL} shows mean tangential fields determined in this way,
plotted as as a function of Galactic longitude. We emphasize that the pulsar
samples used to compute these mean fields are uniformly selected according
to above criteria and that they are independent of any model for the
large-scale structure.

Also plotted in Figure~\ref{B_GL} is the mean line-of-sight field
strength from a simplified model of a bisymmetric spiral field of pitch angle
$-11\degr$ which is counterclockwise within spiral arms and clockwise
in the interarm regions. A rectangular field variation with
Galactocentric radius was assumed, with discontinuous changes in field
direction at the arm-interarm boundaries.  The arm width was assumed
to be equal to interarm width.  Both arm and inter arm fields were
assumed to vary according to Equation~\ref{BR} with a scale radius of
8.5 kpc and a strength at $R=8.5$ kpc of 2.1 $\mu$G.  For each
longitude, the mean field was computed over a path centered on the
tangential point with endpoints defined as described above, thereby
modelling the procedure used to compute the observed mean fields.

While there remains considerable uncertainty in many of the derived field
strengths, overall there is very good agreement between the field directions
predicted by the bisymmetric model and those from the data, giving strong
support to this model for the large-scale field in the Galaxy. The only
places where there is substantial disagreement between the observed and
modelled fields are the Carina region around longitude $280\degr$ and the
Scutum -- Sagittarius interarm region around $l=45\degr$.  As discussed in
\S\ref{car-sag}, mean field strengths over most of the Carina arm are in
accordance with the model, but there is a region of reversed field around
the tangential point -- the Carina anomaly -- which dominates the fits shown
in Figure~\ref{B_GL} for this region. The apparently discrepant point near
$l=45\degr$ can be accounted for by an inward shift of the Sagittarius arm
compared to the equiangular spiral model as discussed in \S\ref{car-sag}.
\begin{figure}[thb]
\includegraphics[angle=270,width=89mm]{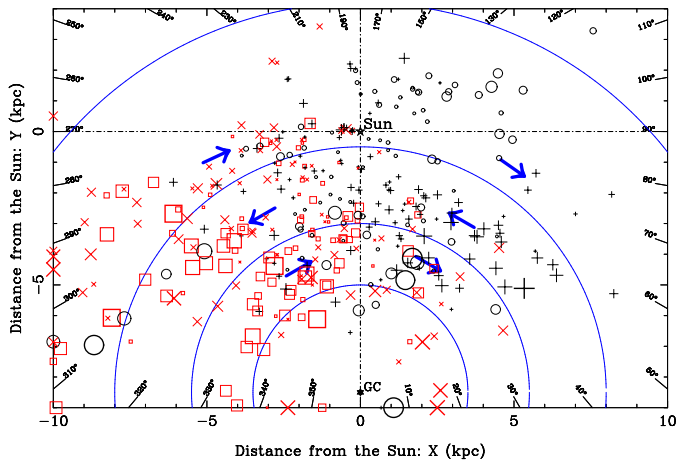}
\caption{The pulsar RM distribution (cf. Figure~\ref{rm-xy}) with
  field directions according to the concentric ring models of
  \citet{rk89}, \citet{rl94} and \citet{val05}.}
\label{rings}
\end{figure}

The model of large-scale fields following the spiral arms is inconsistent
with ring models proposed by \citet{rk89}, \citet{rl94} and \citet{val05}
for RMs of pulsars, which interpret the field structure as azimuthal within
concentric rings centered on the Galactic Center as one option tried by 
\citet{sk80} for RMs of extragalactic radio sources.  Figure~\ref{rings}
illustrates the models of \citet{rk89} and \citet{rl94} with clockwise
fields between Galactocentric radii of 3.5 and 5.5 kpc and between 8.0 and
12.5 kpc. In the middle ring between 5.5 and 8.0 kpc the fields are
counterclockwise. Based on a statistical analysis of previously published
pulsar RM data, \citet{val05} has recently proposed a very similar model
with counterclockwise fields between 5 and 7 kpc and clockwise fields
elsewhere.

These ring models are obviously dominated by the postive RMs of pulsars in
the Sagittarius arm and the negative RMs of pulsars in the Crux
arm. While they were reasonably consistent with earlier observations
in the inner and outer rings, they are inconsistent with the new
observations, especially for the inner ring around longitudes of
$330\degr$ (Norma arm). The  regular reversals of field direction 
from the arm to interarm regions suggested by Figure~\ref{B_GL} are also
inconsistent with the axisymmetric field model of \citet{val96}.

Previous bisymmetric models for the large-scale structure
\citep[e.g.,][]{hq94,id98} have considered configurations where the field
has opposite sign in alternate spiral arms, and consequently half as many
reversals as a function of Galactocentric radius as the model presented
here. These models were primarily based on RMs for pulsars in the local
region, fine within $\sim 3$ kpc of the Sun. With the new RM data set now
available, it is clear that this type of model does not fit the data in 
a larger region.

While there are many caveats regarding pulsar distances and the
locations of spiral arms in the Galaxy, we believe that the evidence
presented here is strongly suggestive that the large-scale magnetic
field in our Galaxy has a bisymmetric form with counterclockwise
fields in the spiral arms and clockwise fields in the interarm
regions. Even with our increased sample, field strengths are poorly
determined in many regions and a larger number of pulsar RMs would
obviously help. The NE2001 pulsar distance model is based on a large
suite of measurements and is believed to be statistically accurate to
5 or 10\% in most directions \citep{cl02}. Individual distances may be
in error by much more than this but, as discussed above, the effect of
such errors is minimised by the averaging procedure used to determine
mean field strengths and directions. The effect of uncertainties in
the locations of spiral arms is also minimised by our emphasis on
field measurements around the tangential points. However, it does
affect the interpretation in interarm regions to a greater extent. One
could reverse the problem and argue that the observed continuity of
field directions over large sections of the assumed spiral arms is
evidence for their reality.

Streaming motion in the Galactic disk associated with the spiral structure
\citep[e.g.][]{sb66} provides a simple mechanism for producing fields of
opposite sign in the arm and interarm regions. In classical density-wave
streaming \citep[e.g.][]{nei92}, peak tangential velocities are against the
direction of rotation on the inner edge of the arm (causing a pile-up of gas
in the arm) and with the direction of rotation on the outer edge of the arm
(resulting in low interarm densities) \citep[e.g.,][]{roh77,aw96}. A
frozen-in, initially radial, field will therefore be aligned with the spiral
pattern with oppositely directed fields in the arm and interarm regions. For
our Galaxy, an initially inwardly directed radial field will become
clockwise in the interarm regions and counterclockwise in the arms. This
idea appears related to the simulations of \citet{gc04a} who find that an
initially azimuthal field has reversed pitch angles in the arm and interarm
regions. Spiral arms are narrower than the interarm regions
\citep[e.g.,][]{aw96,cg02}. We have shown in Figure~\ref{B_R} that interarm 
field strengths are
less than those in spiral arms. Therefore in this model it is quite possible
for the Galaxy as a whole to have zero net magnetic flux.

Even if the Galaxy has zero net magnetic flux, RMs for extragalactic sources
will tend to be dominated by the spiral arm fields because of the
combination of stronger fields and higher electron densities in the arms.
Especially where the path crosses several arms, the effect of the reversed
fields in the interarm regions will not be obvious. Only pulsars with their
distance discrimination can clearly show these reversed fields. Similarly,
if external galaxies have the same field structure as we propose for our
Galaxy, RM measurements are likely to be dominated by the spiral-arm fields
when the resolution is not high enough, leading to the erronous conclusion
that the field is axisymmetric \citep[e.g.,][]{kb98}.

\section{Conclusions}

Pulsars have unique advantages as probes of the large-scale Galactic
magnetic field. Their distribution throughout the Galaxy at
approximately known distances allows a true three-dimensional mapping
of the large-scale field structure. Furthermore, combined with the
measured DMs, pulsar RMs give us a direct measure of the mean
line-of-sight field strength along the path, weighted by the local
electron density. 

We have measured RMs for 223 pulsars, most of which lie in the fourth
and first Galactic quadrants and are relatively distant. These new
measurements enable us to investigate the structure of the Galactic
magnetic field over a much larger region than was previously
possible. Clear evidence is found for coherent large-scale fields
aligned with the spiral-arm structure of the Galaxy. In all of the
inner arms (Norma, Crux-Scutum, Carina-Sagittarius) there is strong
evidence that the large-scale fields are counterclockwise when viewed
from the north Galactic pole. Weaker evidence also suggests that a
counterclockwise field is present in the Perseus arm. On the other
hand, at least in the local region and in the inner Galaxy in the
fourth quadrant, there is good evidence that the fields in interarm
regions are similarly coherent, but clockwise in orientation. Evidence
is also presented that large-scale magnetic fields are stronger in the
inner part of the Galaxy and that fields in interarm regions are
weaker than those in spiral arms. 

We therefore propose a bisymmetric model for the large-scale Galactic
magnetic field with reversals on arm-interarm boundaries so that all
arm fields are counterclockwise and all interarm fields are clockwise.
This model for the Galactic magnetic field is appealing in its
simplicity. It receives strong support from an objective analysis of
mean line-of-sight fields near tangential points of an equiangular
spiral.  However, it clearly needs to be backed up by further
observational work and by modelling of the effects of streaming
motions on various initial field configurations. It is possible that
the mode of initial field formation is not critical in determining the
dominant present-day structure. Further rotation measure observations,
especially for interarm regions and especially in the first Galactic
quadrant, would be especially valuable. 

\section*{Acknowledgments}

We thank the anonymous referee for constructive comments.
JLH is supported by the National Natural Science Foundation (NNSF) of
China (1025313, 10473015, 10521001) and the National Key Basic Research
Science Foundation of China (G19990756). GJQ also thanks the NNSF of China
for support of his visits at ATNF. JLH and GJQ also thank the ATNF
and RNM for their hospitality.  The Parkes telescope is part of the
Australia Telescope which is funded by the Commonwealth Government for
operation as a National Facility managed by CSIRO.

\bibliographystyle{apj}
\bibliography{journals,modrefs,psrrefs,crossrefs}

\newpage \twocolumn

\begin{deluxetable}{clrrrrrrrc}
\tabletypesize{\footnotesize}
\tablecolumns{10}
\tablewidth{0pt}
\tablecaption{Rotation measures for 223 pulsars}
\tablehead{
\colhead{PSR J}  &  \colhead{PSR B} & \colhead{Period} 
& \colhead{DM} & \colhead{$l$}  & \colhead{$b$}   & 
\colhead{Distance} & \colhead{RM} & \colhead{$\sigma_{RM}$} &
\colhead{Obs.} \\
 & & \colhead{(ms)} & \colhead{(cm$^{-3}$ pc)} & \colhead{($\degr$)} &
\colhead{($\degr$)} & \colhead{(kpc)} & \colhead{(rad m$^{-2}$)}& 
\colhead{(rad m$^{-2}$)} & \colhead{Date}
}
\startdata
J0108$-$1431&          &   807.57&     2.38& 140.93& $-$76.82&  0.20& $-$0.3&	  1& Feb03 \\
J0211$-$8159&          &  1077.33&    24.36& 299.59& $-$34.60&  1.00&     54&	  9& Dec00 \\
J0255$-$5304&B0254$-$53&   447.71&    15.90& 269.86& $-$55.31&  0.73&  $-$35&	  3& Dec00 \\
J0421$-$0345&          &  2161.31&    44.61& 197.50& $-$34.75&  2.96&     30&	 14& Dec00 \\
J0448$-$2749&          &   450.45&    26.22& 228.34& $-$37.91&  1.29&     24&	 17& Dec99 \\
J0613$-$0200&          &     3.06&    38.78& 210.41&  $-$9.30&  1.71&     19&	 14& Dec00 \\
J0630$-$2834&B0628$-$28&  1244.42&    34.47& 236.95& $-$16.75&  1.45&     47&	  1& Dec99 \\
J0631$+$1036&          &   287.77&   125.50& 201.21&     0.45&  3.67&    137&	  8& Dec00 \\
J0711$-$6830&          &     5.49&    18.41& 279.53& $-$23.28&  0.86&     41&	  9& Dec00 \\
J0725$-$1635&          &   424.31&    98.98& 231.47&  $-$0.33&  3.68&    188&	 12& Dec99 \\
J0729$-$1448&          &   251.66&    92.30& 230.39&     1.42&  3.52&     55&	  6& Dec00 \\
J0738$-$4042&B0736$-$40&   374.92&   160.80& 254.19&  $-$9.19&  2.64&     11&	  2& Dec00 \\
J0749$-$4247&          &  1095.45&   104.59& 257.06&  $-$8.34&  0.25&     80&	 30& Dec00 \\
J0809$-$4753&B0808$-$47&   547.20&   228.30& 263.30&  $-$7.95&  0.27&    105&	  5& Dec00 \\
J0831$-$4406&          &   311.67&   254.00& 262.28&  $-$2.69&  0.47&    509&	 20& Dec99 \\
J0834$-$4159&          &   121.12&   240.50& 260.88&  $-$1.04&  1.66& $-$377&	 31& Dec00 \\
J0835$-$3707&          &   541.40&   112.30& 257.07&     1.99&  0.60&     68&	 47& Dec00 \\
J0855$-$4644&          &    64.69&   238.20& 266.96&  $-$1.00&  3.88&    249&	 22& Dec00 \\
J0855$-$4658&          &   575.07&   472.70& 267.11&  $-$1.19& 12.81&    350&	 50& Dec00 \\
J0901$-$4624&          &   441.99&   198.80& 267.40&     0.00&  2.82&    289&	 22& Dec00 \\
J0905$-$4536&          &   988.28&   116.80& 267.23&     1.01&  0.62&    153&	 22& Dec99 \\
J0905$-$5127&          &   346.29&   196.43& 271.63&  $-$2.85&  3.29&    292&	  3& Dec99 \\
J0922$-$4949&          &   950.29&   237.10& 272.23&     0.16&  4.17&  $-$15&	  5& Dec99 \\
J0924$-$5814&B0923$-$58&   739.50&    57.40& 278.40&  $-$5.96&  1.86&  $-$45&	  1& Feb03 \\
J0940$-$5428&          &    87.54&   134.50& 277.51&  $-$1.29&  2.95&  $-$18&	 14& Dec00 \\
J0941$-$5244&          &   658.56&   157.94& 276.44&     0.09&  3.14& $-$243&	  4& Dec99 \\
J0954$-$5430&          &   472.83&   200.30& 278.99&  $-$0.10&  3.94&     65&	 10& Dec99 \\
J0959$-$4809&B0957$-$47&   670.09&    92.70& 275.74&     5.41&  2.76&     50&	  6& Dec00 \\
J1000$-$5149&          &   255.68&    72.80& 278.10&     2.60&  1.93&     46&	  9& Dec00 \\
J1001$-$5507&B0959$-$54&  1436.58&   130.00& 280.22&     0.08&  2.77&    297&	 18& Dec00 \\
J1001$-$5559&          &  1661.18&   159.30& 280.69&  $-$0.64&  3.32&    112&	 11& Dec00 \\
J1003$-$4747&B1001$-$47&   307.07&    98.10& 276.04&  $-$6.12&  2.93&     18&	  4& Feb03 \\
J1012$-$5857&B1011$-$58&   819.91&   383.90& 283.70&  $-$2.14&  7.93&     74&	  6& Dec00 \\
J1013$-$5934&          &   442.90&   379.78& 284.13&  $-$2.59&  8.53&  $-$97&	  7& Dec99 \\
J1015$-$5719&          &   139.88&   278.70& 283.08&  $-$0.57&  5.06&    125&	  7& Dec00 \\
J1016$-$5345&B1014$-$53&   769.58&    66.80& 281.20&     2.45&  1.93&  $-$21&	  4& Feb03 \\
J1016$-$5857&          &   107.38&   394.20& 284.07&  $-$1.88&  8.00& $-$537&	 17& Dec00 \\
J1019$-$5749&          &   162.50&  1039.40& 283.83&  $-$0.67&  6.94& $-$366&	 10& Dec00 \\
J1020$-$5921&          &  1238.30&    80.00& 284.71&  $-$1.94&  2.09&  $-$60&	 14& Dec99 \\
J1043$-$6116&          &   288.60&   449.20& 288.22&  $-$2.10&  9.46&    257&	 23& Dec99 \\
J1045$-$4509&          &     7.47&    58.15& 280.85&    12.25&  1.96&     83&	 18& Dec99 \\
J1046$-$5813&B1044$-$57&   369.43&   125.20& 287.07&     0.73&  4.37&    125&	 10& Feb03 \\
J1047$-$6709&          &   198.45&   116.16& 291.31&  $-$7.13&  2.88&  $-$73&	  3& Dec99 \\
J1049$-$5833&          &  2202.33&   446.80& 287.62&     0.64&  7.64&    359&	 11& Dec99 \\
J1054$-$5943&          &   346.91&   330.70& 288.72&  $-$0.10&  5.78&     46&	 34& Dec99 \\
J1103$-$6025&          &   396.59&   275.90& 289.99&  $-$0.29&  4.99&    569&	  7& Dec00 \\
J1110$-$5637&B1107$-$56&   558.25&   262.56& 289.27&     3.53&  5.62&    426&	 11& Dec99 \\
J1112$-$6103&          &    64.96&   599.10& 291.22&  $-$0.46& 12.24&    242&	 15& Dec00 \\
J1112$-$6613&B1110$-$65&   334.21&   249.30& 293.19&  $-$5.23&  6.48&  $-$94&	 18& Dec99 \\
J1116$-$4122&B1114$-$41&   943.16&    40.53& 284.45&    18.06&  1.47&  $-$37&	 13& Dec00 \\
J1117$-$6154&          &   505.10&   493.60& 292.10&  $-$1.02&  8.91& $-$622&	 10& Dec99 \\
J1119$-$6127&          &   407.75&   707.40& 292.15&  $-$0.53& 17.14&    832&	  6& Dec00 \\
J1121$-$5444&B1119$-$54&   535.78&   204.70& 290.08&     5.87&  5.17&     42&	  9& Feb03 \\
J1123$-$6102&          &   640.23&   439.40& 292.50&     0.04&  7.85&    244&	  8& Dec99 \\
J1123$-$6259&          &   271.43&   223.26& 293.18&  $-$1.78&  4.28&     54&	 10& Dec99 \\
J1124$-$6421&          &   479.10&   298.00& 293.74&  $-$3.03&  6.66&$-$1102&	 98& Dec00 \\
J1126$-$6054&B1124$-$60&   202.74&   280.27& 292.83&     0.29&  5.31&  $-$41&	 16& Dec00 \\
J1133$-$6250&B1131$-$62&  1022.87&   567.80& 294.21&  $-$1.29& 12.10&    880&	 24& Dec00 \\
J1137$-$6700&          &   556.22&   228.04& 295.79&  $-$5.16&  5.70&   $-$1&	 13& Dec99 \\
J1138$-$6207&          &   117.56&   519.80& 294.50&  $-$0.46&  9.65&    594&	 18& Dec00 \\
J1141$-$3107&          &   538.43&    30.77& 285.74&    29.39&  1.21&  $-$60&	 30& Dec00 \\
J1141$-$3322&          &   291.47&    46.45& 286.58&    27.27&  1.91&  $-$33&	 14& Dec00 \\
J1141$-$6545&          &   393.90&   116.05& 295.79&  $-$3.86&  2.47&     84&	  2& Feb03 \\
J1146$-$6030&B1143$-$60&   273.37&   112.80& 294.97&     1.34&  2.35&     10&	 17& Dec00 \\
J1157$-$6224&B1154$-$62&   400.52&   325.20& 296.70&  $-$0.19&  6.31&    510&	  2& Dec00 \\
J1159$-$6409&          &   667.49&   178.00& 297.29&  $-$1.86&  3.41&    259&	 13& Dec00 \\
J1159$-$7910&          &   525.07&    59.24& 300.41& $-$16.55&  1.91&  $-$11&	  9& Dec99 \\
J1211$-$6324&          &   433.08&   333.80& 298.47&  $-$0.88&  6.96& $-$120&	  9& Dec00 \\
J1224$-$6208&          &   585.76&   454.20& 299.81&     0.56& 10.01&     80&	  7& Dec00 \\
J1224$-$6407&B1221$-$63&   216.48&    96.90& 299.98&  $-$1.41&  3.14&      6&	  3& Dec99 \\
J1225$-$6408&B1222$-$63&   419.62&   416.00& 300.13&  $-$1.41& 10.46&    356&	 23& Dec00 \\
J1231$-$6303&          &  1351.24&   301.00& 300.64&  $-$0.27&  6.88&    390&	 18& Dec00 \\
J1239$-$6832&B1236$-$68&  1301.92&    94.30& 301.88&  $-$5.69&  2.15&  $-$49&	 10& Feb03 \\
J1252$-$6314&          &   823.34&   278.40& 303.07&  $-$0.37&  5.27&    330&	 19& Dec99 \\
J1301$-$6305&          &   184.53&   374.00& 304.10&  $-$0.24&  6.65& $-$665&	 31& Dec00 \\
J1305$-$6203&          &   427.76&   470.00& 304.56&     0.77&  8.51& $-$436&	 15& Dec00 \\
J1306$-$6617&B1303$-$66&   473.03&   436.90& 304.46&  $-$3.46& 12.38&    387&	 10& Dec00 \\
J1307$-$6318&          &  4962.43&   374.00& 304.78&  $-$0.49&  6.62&    136&	 23& Dec00 \\
J1312$-$6400&          &  2437.43&    93.00& 305.19&  $-$1.23&  1.94&     40&	 30& Dec00 \\
J1312$-$5402&B1309$-$53&   728.15&   133.00& 305.99&     8.70&  3.42&    143&	 20& Feb03 \\
J1317$-$6302&          &   261.27&   678.10& 305.90&  $-$0.32& 12.06& $-$504&	  8& Dec00 \\
J1322$-$6241&          &   506.06&   618.80& 306.48&  $-$0.04& 10.23&     83&	 14& Dec99 \\
J1324$-$6146&          &   844.11&   828.00& 306.85&     0.85& 10.12&$-$1546&	 29& Dec99 \\
J1326$-$6700&B1322$-$66&   543.01&   209.60& 306.31&  $-$4.37&  4.81&  $-$47&	  1& Feb03 \\
J1327$-$6222&B1323$-$62&   529.91&   318.40& 307.07&     0.20&  5.50& $-$306&	  8& Dec99 \\
J1327$-$6301&B1323$-$62&   196.48&   294.91& 306.96&  $-$0.42&  5.26&     96&	 12& Dec00 \\
J1327$-$6400&          &   280.68&   680.90& 306.83&  $-$1.40& 15.49& $-$141&	 58& Dec00 \\
J1334$-$5839&          &   107.72&   119.30& 308.52&     3.74&  2.41&     54&	  7& Dec99 \\
J1341$-$6023&          &   627.29&   364.60& 309.03&     1.88&  7.04& $-$688&	 23& Dec00 \\
J1345$-$6115&          &  1253.08&   278.10& 309.41&     0.92&  4.98&  $-$61&	  2& Dec99 \\
J1347$-$5947&          &   609.96&   293.40& 309.91&     2.31&  5.82& $-$548&	  6& Dec00 \\
J1349$-$6130&          &   259.36&   284.60& 309.81&     0.58&  4.98& $-$380&	 13& Dec99 \\
J1350$-$5115&          &   295.70&    90.39& 312.23&    10.54&  2.28&   $-$2&	 22& Dec99 \\
J1355$-$5153&B1352$-$51&   644.30&   112.10& 312.95&     9.71&  2.83&  $-$55&	  6& Feb03 \\
J1355$-$6206&          &   276.60&   547.00& 310.33&  $-$0.15&  8.28& $-$474&	  6& Dec00 \\
J1356$-$5521&          &   507.38&   174.17& 312.19&     6.33&  4.19&    150&	 22& Dec00 \\
J1356$-$6230&B1353$-$62&   455.77&   417.30& 310.41&  $-$0.58&  6.61& $-$586&	  5& Dec99 \\
J1357$-$6429&          &   166.07&   129.30& 309.92&  $-$2.51&  2.51&  $-$54&	 13& Dec00 \\
J1403$-$6310&          &   399.17&   305.00& 310.92&  $-$1.42&  5.44& $-$709&	 21& Dec00 \\
J1406$-$5806&          &   288.35&   229.00& 312.67&     3.35&  4.79&    153&	  4& Feb03 \\
J1406$-$6121&          &   213.07&   542.30& 311.84&     0.20&  8.15&    880&	 60& Dec00 \\
J1412$-$6145&          &   315.23&   514.70& 312.32&  $-$0.36&  7.82&  $-$39&	 20& Dec00 \\
J1413$-$6141&          &   285.62&   677.00& 312.46&  $-$0.33& 10.14&  $-$35&	 10& Dec00 \\
J1413$-$6222&          &   292.41&   808.10& 312.24&  $-$0.98& 15.51& $-$490&	  8& Dec99 \\
J1413$-$6307&          &   394.95&   121.98& 312.05&  $-$1.71&  2.34&     45&	  9& Dec99 \\
J1416$-$6037&          &   295.58&   289.20& 313.17&     0.53&  4.84&    336&	  7& Dec99 \\
J1420$-$5416&B1417$-$54&   935.77&   129.60& 315.79&     6.36&  2.88&   $-$1&	  9& Feb03 \\
J1420$-$6048&          &    68.18&   360.00& 313.54&     0.22&  5.63& $-$110&	 16& Dec00 \\
J1424$-$5822&          &   366.73&   323.90& 314.89&     2.31&  6.21& $-$625&	 19& Dec99 \\
J1452$-$5851&          &   386.62&   262.40& 318.08&     0.39&  4.30&     47&	  7& Dec00 \\
J1452$-$6036&          &   154.99&   349.70& 317.29&  $-$1.16&  5.79&     10&	  5& Dec99 \\
J1502$-$5828&          &   668.11&   584.00& 319.40&     0.13&  8.16&    362&	  7& Feb03 \\
J1507$-$4352&B1504$-$43&   286.76&    48.70& 327.33&    12.45&  1.34&  $-$33&	  4& Dec99 \\
J1510$-$4422&B1507$-$44&   943.87&    84.00& 327.59&    11.73&  2.14&      8&	  8& Feb03 \\
J1512$-$5759&B1508$-$57&   128.69&   628.70& 320.77&  $-$0.10&  7.35&    513&	 16& Dec99 \\
J1513$-$5739&          &   973.46&   469.70& 321.09&     0.10&  6.45&    261&	 13& Dec99 \\
J1530$-$5327&          &   278.96&    49.60& 325.32&     2.34&  1.23&  $-$19&	 21& Dec99 \\
J1531$-$5610&          &    84.20&   110.90& 323.89&     0.03&  2.09&  $-$50&	 20& Dec00 \\
J1534$-$5405&B1530$-$53&   289.69&   190.82& 325.46&     1.48&  3.37&  $-$46&	 17& Dec00 \\
J1536$-$5433&          &   881.44&   147.50& 325.37&     0.98&  2.71& $-$155&	 13& Dec99 \\
J1540$-$5736&          &   612.92&   304.50& 324.10&  $-$1.89&  5.07& $-$414&	 15& Dec00 \\
J1541$-$5535&          &   295.84&   428.00& 325.42&  $-$0.33&  5.74& $-$256&	 13& Dec00 \\
J1543$-$5459&          &   377.12&   345.70& 326.02&  $-$0.04&  4.84&     28&	 23& Dec00 \\
J1546$-$5302&          &   580.84&   287.00& 327.47&     1.30&  5.20&$-$1135&	 90& Dec00 \\
J1548$-$5607&          &   170.93&   315.50& 325.85&  $-$1.35&  4.86&     37&	 10& Dec00 \\
J1550$-$5242&          &   749.66&   337.70& 328.14&     1.19&  6.63& $-$440&	 25& Dec00 \\
J1556$-$5358&          &   994.68&   436.00& 328.11&  $-$0.43&  6.31& $-$153&	 16& Dec99 \\
J1600$-$5751&B1556$-$57&   194.45&   176.55& 325.97&  $-$3.70&  3.49& $-$131&	  8& Feb03 \\
J1601$-$5335&          &   288.46&   194.60& 328.93&  $-$0.62&  4.55& $-$157&	 35& Dec00 \\
J1603$-$2531&          &   283.07&    53.76& 348.37&    19.98&  1.87&     15&	  4& Dec00 \\
J1603$-$7202&          &    14.84&    38.05& 316.63& $-$14.49&  1.17&     30&	  6& Dec00 \\
J1610$-$5006&          &   481.12&   416.00& 332.27&     1.05&  7.54& $-$756&	 23& Dec99 \\
J1610$-$5303&          &   786.47&   380.10& 330.21&  $-$1.06&  6.74& $-$335&	 30& Dec00 \\
J1611$-$4949&          &   666.44&   556.80& 332.59&     1.14&  9.52& $-$405&	 22& Dec99 \\
J1611$-$5209&B1607$-$52&   182.49&   128.20& 330.92&  $-$0.48&  4.35&  $-$72&	  6& Dec99 \\
J1613$-$4714&B1609$-$47&   382.38&   161.20& 334.57&     2.84&  3.69& $-$138&	  7& Feb03 \\
J1614$-$5048&B1610$-$50&   231.69&   582.80& 332.21&     0.17&  7.94& $-$451&	  2& Feb03 \\
J1615$-$5444&          &   360.96&   312.60& 329.57&  $-$2.76&  5.70& $-$232&	 28& Dec00 \\
J1618$-$4723&          &   203.55&   134.70& 335.04&     2.18&  2.99&     39&	  4& Feb03 \\
J1623$-$2631&B1620$-$26&    11.08&    62.86& 350.97&    15.96&  1.80&   $-$8&	 20& Dec00 \\
J1623$-$4256&B1620$-$42&   364.59&   295.00& 338.89&     4.62&  6.58&  $-$15&	  8& Feb03 \\
J1623$-$4949&          &   725.73&   183.30& 334.00&  $-$0.21&  3.57&  $-$42&	  7& Dec00 \\
J1625$-$4048&          &  2355.28&   145.00& 340.60&     5.93&  3.14&   $-$7&	 15& Dec00 \\
J1628$-$4804&          &   865.97&   952.00& 335.76&     0.46&  9.79& $-$431&	 43& Dec00 \\
J1630$-$4719&          &   559.07&   489.60& 336.49&     0.78&  5.79& $-$339&	 10& Dec00 \\
J1630$-$4733&B1626$-$47&   575.97&   498.00& 336.40&     0.56&  5.65& $-$338&	  8& Dec00 \\
J1632$-$4818&          &   813.45&   758.00& 336.08&  $-$0.20&  7.77& $-$515&	 39& Dec00 \\
J1633$-$4453&B1630$-$44&   436.51&   475.40& 338.72&     1.98&  7.12&    139&	 17& Dec00 \\
J1635$-$4944&          &   671.96&   474.00& 335.39&  $-$1.57&  6.62&  $-$23&	 15& Dec00 \\
J1637$-$4553&B1634$-$45&   118.77&   193.23& 338.47&     0.76&  3.16&     12&	  4& Dec99 \\
J1637$-$4642&          &   154.03&   417.00& 337.78&     0.31&  5.08&     13&	 18& Dec00 \\
J1638$-$4608&          &   278.14&   424.30& 338.34&     0.54&  5.18&    335&	 12& Dec00 \\
J1639$-$4359&          &   587.56&   258.90& 340.02&     1.87&  4.02&    129&	 18& Dec00 \\
J1639$-$4604&B1635$-$45&   264.56&   259.00& 338.50&     0.45&  3.76&  $-$60&	 30& Dec00 \\
J1640$-$4715&B1636$-$47&   517.40&   592.00& 337.71&  $-$0.43&  6.48& $-$398&	 22& Dec99 \\
J1643$-$1224&          &     4.62&    62.41&   5.66&    21.21&  2.41& $-$263&	 15& Dec00 \\
J1644$-$4559&B1641$-$45&   455.06&   478.80& 339.19&  $-$0.20&  5.09& $-$617&	  1& Feb03 \\
J1646$-$4346&B1643$-$43&   231.60&   490.00& 341.10&     0.96&  5.79&  $-$62&	  7& Dec00 \\
J1648$-$4611&          &   164.95&   392.90& 339.43&  $-$0.79&  4.96& $-$682&	 26& Dec00 \\
J1649$-$4349&          &   870.71&   398.60& 341.36&     0.59&  5.02&    759&	 17& Dec00 \\
J1650$-$4502&          &   380.87&   319.70& 340.55&  $-$0.35&  4.42&    130&	 10& Dec00 \\
J1651$-$4246&B1648$-$42&   844.08&   525.00& 342.45&     0.92&  6.35& $-$154&	  5& Dec00 \\
J1653$-$3838&B1650$-$38&   305.04&   207.20& 345.87&     3.26&  3.66&  $-$74&	  6& Dec99 \\
J1653$-$4249&          &   612.56&   416.10& 342.63&     0.62&  5.24&     25&	 17& Dec99 \\
J1701$-$3726&B1658$-$37&  2454.61&   303.40& 347.75&     2.83&  5.17& $-$602&	  8& Dec99 \\
J1701$-$4533&B1657$-$45&   322.91&   526.00& 341.36&  $-$2.17&  9.71&     17&	 13& Dec00 \\
J1703$-$4851&          &  1396.40&   150.29& 338.98&  $-$4.51&  2.99&   $-$4&	 24& Dec00 \\
J1705$-$3423&          &   255.43&   146.36& 350.72&     3.97&  2.85&  $-$44&	  8& Dec00 \\
J1705$-$4108&          &   861.07&  1077.00& 345.29&  $-$0.04& 11.43&    916&	 15& Dec99 \\
J1708$-$3426&          &   692.11&   190.70& 351.08&     3.40&  3.56& $-$176&	 15& Dec00 \\
J1715$-$3903&          &   278.48&   313.10& 348.10&  $-$0.32&  4.11&    250&	 15& Dec00 \\
J1717$-$3425&B1714$-$34&   656.30&   587.70& 352.12&     2.02&  9.98& $-$191&	 14& Dec00 \\
J1717$-$4043&          &   397.86&   452.60& 347.01&  $-$1.69&  6.28& $-$993&	 17& Dec99 \\
J1718$-$3825&          &    74.67&   247.40& 348.95&  $-$0.43&  3.60&    113&	 10& Dec00 \\
J1719$-$4006&B1715$-$40&   189.09&   386.60& 347.65&  $-$1.53&  5.13& $-$234&	 31& Dec00 \\
J1720$-$1633&B1717$-$16&  1565.60&    44.83&   7.37&    11.54&  1.34&  $-$12&	 13& Feb03 \\
J1720$-$3659&          &   351.12&   381.60& 350.33&     0.10&  4.59&  $-$99&	  7& Dec00 \\
J1722$-$3632&B1718$-$36&   399.18&   416.20& 350.93&     0.00&  4.35& $-$307&	  8& Dec99 \\
J1723$-$3659&          &   202.72&   254.20& 350.68&  $-$0.40&  3.54& $-$219&	 11& Dec00 \\
J1730$-$3350&B1727$-$33&   139.46&   258.00& 354.13&     0.09&  3.54& $-$132&	 10& Dec00 \\
J1733$-$3716&B1730$-$37&   337.59&   155.10& 351.57&  $-$2.28&  2.80& $-$330&	  6& Dec00 \\
J1738$-$3211&B1735$-$32&   768.50&    49.59& 356.46&  $-$0.49&  1.20&      7&	  9& Dec00 \\
J1739$-$2903&B1736$-$29&   322.88&   138.56& 359.20&     1.06&  2.47& $-$236&	 18& Dec00 \\
J1739$-$3023&          &   114.37&   170.00& 358.08&     0.33&  2.91&  $-$74&	 18& Dec00 \\
J1741$-$3927&B1737$-$39&   512.21&   158.50& 350.55&  $-$4.74&  3.21&    204&	  6& Dec00 \\
J1743$-$3150&B1740$-$31&  2414.58&   193.05& 357.29&  $-$1.14&  3.31& $-$240&	 12& Dec00 \\
J1750$-$3157&B1747$-$31&   910.36&   206.34& 357.98&  $-$2.51&  3.82&    111&	  8& Dec00 \\
J1757$-$2421&B1754$-$24&   234.10&   179.44&   5.28&     0.05&  4.40&   $-$9&	  9& Dec99 \\
J1759$-$2205&B1756$-$22&   460.97&   177.16&   7.47&     0.81&  3.57&      1&	 10& Dec99 \\
J1801$-$2451&B1757$-$24&   124.92&   289.00&   5.25&  $-$0.88&  5.22&    637&	 12& Dec00 \\
J1808$-$2057&B1805$-$20&   918.41&   606.80&   9.44&  $-$0.40&  7.61&     93&	 11& Dec00 \\
J1809$-$1917&          &    82.75&   197.10&  11.09&     0.08&  3.55&    130&	 12& Dec00 \\
J1809$-$2109&B1806$-$21&   702.41&   381.91&   9.41&  $-$0.72&  5.23&    256&	 24& Dec00 \\
J1809$-$3547&          &   860.39&   193.84& 356.54&  $-$7.76&  5.39&    379&	 18& Dec99 \\
J1817$-$3837&          &   384.49&   102.85& 354.67& $-$10.40&  2.51&     89&	  6& Dec99 \\
J1818$-$1422&B1815$-$14&   291.49&   622.00&  16.40&     0.61&  7.15&   1168&	 13& Dec00 \\
J1818$-$1519&          &   939.69&   845.00&  15.55&     0.19&  9.55&   1157&	 23& Dec00 \\
J1820$-$1346&B1817$-$13&   921.46&   776.70&  17.16&     0.48&  8.83&    893&	 12& Dec00 \\
J1820$-$1818&B1817$-$18&   309.90&   436.00&  13.20&  $-$1.72&  7.04&  $-$60&	 24& Dec00 \\
J1822$-$4209&          &   456.51&    72.51& 351.88& $-$12.82&  1.86&  $-$13&	  9& Dec99 \\
J1823$-$0154&          &   759.78&   135.87&  28.08&     5.25&  3.62&    153&	 24& Dec00 \\
J1823$-$1115&B1820$-$11&   279.83&   428.59&  19.76&     0.94&  5.59& $-$354&	 10& Dec00 \\
J1824$-$1118&B1821$-$11&   435.76&   603.00&  19.80&     0.74&  7.09&    213&	 17& Dec00 \\
J1824$-$1159&          &   362.49&   463.40&  19.25&     0.32&  5.56&    407&	 15& Dec00 \\
J1827$-$0934&          &   512.55&   259.20&  21.72&     0.84&  4.19& $-$386&	 24& Dec00 \\
J1835$-$0643&B1832$-$06&   305.83&   472.90&  25.09&     0.55&  6.21&     62&	 38& Dec99 \\
J1845$-$0434&B1842$-$04&   486.75&   230.80&  28.19&  $-$0.79&  5.07& $-$248&	 10& Feb03 \\
J1856$+$0113&B1853$+$01&   267.44&    96.79&  34.56&  $-$0.49&  3.30& $-$140&	 30& Dec00 \\
J1857$+$0057&B1854$+$00&   356.93&    82.39&  34.41&  $-$0.80&  2.79&     79&	 26& Dec00 \\
J1857$+$0212&B1855$+$02&   415.82&   506.77&  35.61&  $-$0.39&  7.98&    423&	 21& Dec00 \\
J1857$+$0943&B1855$+$09&     5.36&    13.31&  42.29&     3.06&  0.91&     53&	  9& Dec00 \\
J1901$+$0156&B1859$+$01&   288.22&   105.39&  35.81&  $-$1.36&  2.79& $-$122&	  9& Dec00 \\
J1904$+$0004&          &   139.52&   233.61&  34.45&  $-$2.81&  5.74&    306&	  9& Dec00 \\
J1915$+$1606&B1913$+$16&    59.03&   168.77&  49.96&     2.12&  5.90&    430&	 73& Dec00 \\
J1946$-$2913&B1943$-$29&   959.45&    44.31&  11.10& $-$24.12&  1.54&  $-$28&	 10& Dec99 \\
J2038$-$3816&          &  1577.29&    33.96&   3.85& $-$36.74&  1.36&     68&	 18& Dec00 \\
J2053$-$7200&B2048$-$72&   341.34&    17.30& 321.87& $-$34.99&  0.48&     15&	  9& Dec99 \\
J2124$-$3358&          &     4.93&     4.62&  10.92& $-$45.43&  0.27&      5&	  2& Dec00 \\
J2129$-$5721&          &     3.73&    31.86& 338.00& $-$43.57&  1.36&     30&	  5& Dec00 \\
J2145$-$0750&          &    16.05&     9.00&  47.77& $-$42.08&  0.57&     12&	  6& Dec00 \\
J2248$-$0101&          &   477.23&    29.05&  69.25& $-$50.62&  1.65&     33&	 12& Dec00 \\
J2324$-$6054&B2321$-$61&  2347.49&    16.00& 320.42& $-$53.17&  0.69&  $-$11&	  8& Dec00 \\
J2330$-$2005&B2327$-$20&  1643.62&     8.46&  49.39& $-$70.19&  0.39&     30&	  7& Dec00 \\
\enddata
\end{deluxetable}

\begin{table*}
\tabletypesize{\footnotesize}
\tablecolumns{7}
\tablewidth{0pt}
\caption{Comparison of derived RMs with previously published values}
\begin{tabular}{llrrcrc}
\hline
\hline
\multicolumn{1}{c}{PSR J} & \multicolumn{1}{c}{PSR B} &
\multicolumn{1}{c}{RM}   &  
\multicolumn{1}{c}{Prev. RM} &  Ref. &  \multicolumn{1}{c}{Prev. RM} & Ref.\\
  & & \multicolumn{1}{c}{(rad m$^{-2}$)} & \multicolumn{1}{c}{(rad m$^{-2}$)} & &
\multicolumn{1}{c}{(rad m$^{-2}$)} \\
\hline
J0108$-$1431&           &   $-$1$\pm$1  &   $-$1$\pm$3   & 1&             &  \\
J0255$-$5304& B0254$-$53&  $-$35$\pm$3  &  $-$10$\pm$12  & 1&             &  \\
J0630$-$2834& B0628$-$28&   46.6$\pm$1.3&   46.2$\pm$0.1 & 2&45.7$\pm$0.5 & 3   \\
J0738$-$4042& B0736$-$40&     11$\pm$2  &   13.5$\pm$0.4 & 2&14.5$\pm$0.7 & 3   \\
J0809$-$4753& B0808$-$47&    105$\pm$5  &    100$\pm$20  & 4&97$\pm$4     & 5   \\
J1112$-$6613& B1110$-$65&  $-$94$\pm$18 & $-$370$\pm$50  & 4&	          &	\\
J1116$-$4122& B1114$-$41&  $-$37$\pm$13 &     55$\pm$40  & 3&  31$\pm$16  & 1 \\
J1119$-$6127&           &    832$\pm$6  &    842$\pm$23  & 6&	      &	       \\
J1141$-$6545&           &     84$\pm$2  &  $-$86$\pm$3   & 7&	      &	       \\
J1146$-$6030& B1143$-$60&     10$\pm$17 &     30$\pm$20  & 4&	      &	       \\
J1157$-$6224& B1154$-$62&    510$\pm$2  &  508.2$\pm$0.5 & 2&         &	       \\
J1224$-$6407& B1221$-$63&      6$\pm$3  &    $3.6\pm$0.5 & 2&         &	       \\
J1420$-$6048&           & $-$110$\pm$16 & $-$106$\pm$18  & 8&	      &	       \\
J1614$-$5048& B1610$-$50& $-$451$\pm$2  & $-$560$\pm$60  & 5&	      &	       \\
J1644$-$4559& B1641$-$45& $-$617$\pm$1  & $-$611$\pm$2   & 3&	      &	       \\
J1646$-$4346& B1643$-$43&  $-$62$\pm$7  &    $-65\pm$17  & 9&	      &	       \\
J1701$-$4533& B1657$-$45&     17$\pm$13 &     60$\pm$30  & 5&	      &	       \\
J1730$-$3350& B1727$-$33& $-$132$\pm$10 &   $-142\pm$05  & 9&	      &	       \\
J1741$-$3927& B1737$-$39&    204$\pm$6  &    221$\pm$29  & 3& 180$\pm$14 & 1 \\
J1757$-$2421& B1754$-$24&   $-$9$\pm$9  &      0$\pm$12$^*$  & 10&	      &	       \\
J1759$-$2205& B1756$-$22&      1$\pm$10 &      6$\pm$7   &10&	      &	       \\
J1857$+$0057& B1854$+$00&     79$\pm$26 &    104$\pm$19  &11&	      &	       \\
J1946$-$2913& B1943$-$29&  $-$28$\pm$10 &      8$\pm$7   & 1&	      &	       \\
J2038$-$3816&           &     68$\pm$18 &     30$\pm$15  & 1&         &	       \\
J2053$-$7200& B2048$-$72&     15$\pm$9  &   17.0$\pm$1.0 & 5& 9$\pm$4 &  1 \\
J2324$-$6054& B2321$-$61&  $-$11$\pm$8  &   39.0$\pm$6.0 & 5& 	      &	       \\
J2330$-$2005& B2327$-$20&     30$\pm$7  &    9.5$\pm$0.2 & 2& 16$\pm$3&  10 \\  
\hline
\hline
\multicolumn{7}{l}{\small \parbox[t]{12cm}{References: 
1: \citet{hmq99}; 2: \citet{hmm81}; 3: \citet{vdhm97}; 4: \citet{cmh91}; 
5: \citet{qmlg95}; 6: \citet{ck03}; 
7: \citet{klm+00};
8: \citet{rrj01}; 9: \cite{cmk01}; 
10: \citet{hl87}; 11: \citet{wck+04}. $^*$ Note: see text.
}}
\end{tabular}
\end{table*}

\end{document}